\documentclass[aps,showpacs,floatfix,superscriptaddress,12pt,prb]{revtex4-1}
\usepackage{amsmath}
\usepackage{latexsym}
\usepackage{geometry}
\usepackage[english]{babel}
\usepackage{blindtext}
\usepackage{graphicx}
\usepackage{subfigure}
\usepackage{indentfirst}
\usepackage{color}
\usepackage[colorlinks,linkcolor=blue,anchorcolor=blue,citecolor=green]{hyperref}

\graphicspath{{figure/}}

\begin{document}

\title{Enhanced, phase coherent, multifractal-like, two-dimensional superconductivity}
\author{Bo Fan}
\email{bo.fan@sjtu.edu.cn}
\affiliation{Shanghai Center for Complex Physics, 
	School of Physics and Astronomy, Shanghai Jiao Tong
	University, Shanghai 200240, China}
\author{Antonio M. Garc\'\i a-Garc\'\i a}
\email{amgg@sjtu.edu.cn}
\affiliation{Shanghai Center for Complex Physics, 
	School of Physics and Astronomy, Shanghai Jiao Tong
	University, Shanghai 200240, China}

\begin{abstract}We study the interplay of superconductivity and disorder by solving numerically the Bogoliubov-de-Gennes equations in a two dimensional lattice of size $80\times80$ which makes possible to investigate the weak-coupling limit. In contrast with results in the strong coupling region, we observe enhancement of superconductivity and intriguing multifractal-like features such as a broad log-normal spatial distribution of the order parameter, a parabolic singularity spectrum, and level statistics consistent with those of a disordered metal at the Anderson transition. The calculation of the superfluid density, including small phase fluctuations, reveals that, despite this intricate spatial structure, phase coherence still holds for sufficiently weak disorder. It only breaks down for stronger disorder but before the insulating transition takes place. 
\end{abstract}
\maketitle

\section{\textbf{Introduction}}

Disorder in low dimensions amplifies wave-like aspects of quantum matter. For instance, Anderson localization, caused by quantum interference, stops classical diffusion in two dimensional disordered systems \cite{Anderson1958,abraham1979}. The interplay between quantum coherence phenomena and interactions is not yet well understood. In the case of superconductivity, it was believed \cite{Anderson1959,Abrikosov1961} for many years that disorder was not important even in low spatial dimensions because in the presence of elastic scattering by impurities Cooper pairs could still have approximately zero net momentum. Furthermore, results from a mean-field Bardeen-Cooper-Schrieffer (BCS) approach \cite{Ma1985,Sadovskii1997} suggested that pairing could survive even in the  insulator phase. However, BCS or phenomenological Ginzburg-Landau approaches, oversimplifies many aspects of the rich phenomenology of quantum coherence effects in interacting systems. Moreover, these quantum effects are observed experimentally mostly in low dimensions and very low temperatures beyond the range that could be reached experimentally at that time.

Indeed, more quantitative numerical studies \cite{ghosal1998,Ghosal2001,Trivedi2012,Dubi2007}, revealed a rather different picture. Even in the limit of strong electron-phonon coupling, disorder was found to cause substantial spatial inhomogeneities and an emergent granularity \cite{Trivedi2012} in the spatial distribution of the superconducting order parameter in two dimensions. Recent experimental results \cite{brun2014,Lemarie2013,noat2013,verdu2018,xue2019,Dubi2007} have indeed confirmed that in low dimensions even a relatively weak disorder can lead to a highly inhomogeneous superconducting state with an intricate spatial structure.

Theoretical studies \cite{cea2014,pracht2017} have also linked the experimental observation of a collective Goldstone mode as a sub-gap excitation with the existence of a highly inhomogeneous superconducting state. The critical temperature at the metal-insulator transition, characterized by spatially multifractal eigenstates \cite{Castellani1986,Wegner1980}, was predicted to be dramatically enhanced with respect to the homogeneous limit in both three \cite{Feigelman2007,Feigelman2010} and two dimensions \cite{Burmistrov2012}. However, such massive enhancement has never been observed experimentally. 
In the weak-disorder and weak-coupling limit, it was found \cite{Mayoh2015,Mayoh2014a,Brian2018} that the superconducting gap has a broad log-normal spatial distribution and that the critical temperature is enhanced modestly but only for sufficiently weak electron-phonon interactions. These results are also consistent with previous reports of enhancement of phase coherence \cite{Tezuka2010} in one dimensional superconductors at zero temperature close to the superconductor-insulator transition.
Experimental results in granular Al \cite{pracht2016,pracht2017,Deutscher1973,Abeles1966} and in one-layer NbSe$_2$ \cite{verdu2018,xue2019}, a weakly coupled superconductor, have indeed confirmed the theoretical prediction of Ref.~\cite{Mayoh2015} with respect to the enhancement of superconductivity and the log-normal distribution of the superconducting gap. 

Despite these advances, we are still far from a full understanding of the effect of disorder in low dimensional superconductors. Analytical results \cite{Feigelman2007,Feigelman2010,Mayoh2015} are mostly based on a BCS approach where the spatial inhomogeneity of the order parameter is directly borrowed from that of the one-body problem thus missing any many-body effects. For instance, the BCS formalism for disordered superconductors is not fully self-consistent. By contrast, the Bogoliubov de-Gennes (BdG) equations, though still a mean-field approach, yields fully self-consistent solutions for the order parameter and therefore it is especially suited to investigate the interplay of disorder and superconductivity.  

Previous numerical findings based on the Bogoliubov de-Gennes equations \cite{Ghosal2001,ghosal1998} in two dimensions, which yields fully self-consistent solutions, are typically carried out in the limit of very strong electron-phonon coupling constant and no Debye energy in order that the system size is much larger than the superconducting coherence length. Quantitative comparisons with experiments are in principle not possible.  

Here, we carry out an extensive numerical investigation of a two-dimensional disordered superconductor by solving numerically the BdG equations and also including the effect of small phase fluctuations. We study comparatively large lattice sizes $80\times 80$ which allows us to employ a finite Debye energy and a relatively weak electron-phonon coupling $|U| \sim 1$. This range of parameters is closer to the one believed to describe many two-dimensional superconducting materials. 
 
Indeed, we found qualitative differences with the strong coupling results of  Ref.~\cite{Ghosal2001,ghosal1998} that we now summarize: 
only in the weak-coupling limit, the spatial distribution of the order parameter is well described by a log-normal distribution. The analysis of the singularity spectrum related to the order parameter spatial distribution is broad and parabolic only in this weak-coupling limit. This is the prediction for a weakly multifractal measure. We have also observed that, also in stark contrast with the strong coupling results, both the spectral gap and spatial average of the order parameter of disordered superconductors are enhanced by disorder. This feature has its origin in that, as disorder increases, the number of strongly overlapping eigenstates of the BdG model close to the Fermi energy grows as well. Level statistics of eigenvalues of the BdG Hamiltonian in the intermediate disorder region is between Wigner-Dyson and Poisson statistics, and strikingly similar to those found in systems \cite{altshuler1988repulsion,shapiro1993,mirlin1996,wang2009,huang2019anderson} at the Anderson metal-insulator transition.    
Likewise, the superfluid density, that includes the effect of phase fluctuations to leading order, is still finite. This is an indication that phase coherence still holds despite strong spatial fluctuations induced by disorder. For sufficiently strong disorder, the superfluid stiffness vanishes but the amplitude of the order parameter is still finite which hints the possibility of an intermediate metallic state before the superconductor insulator transition occurs. 

Our results points to a rich phenomenology of weakly disordered superconductors in low dimensions of potential relevance for many two dimensional materials. It could also shed light on long standing problems in the field such as the existence and characterization of an intermediate metallic state in the low temperature limit, which has been observed in several two and quasi-two dimensional systems \cite{han2014,Eley2011,sajadi2018,bottcher2017,Jaeger1989,Haviland1989,goldman1993,saito2015,ye2012,barriocanal2013,Caviglia2008,tsen2016,breznay2017,saito2015,ephron1996,qin2006}, or the surprising enhancement of superconductivity observed in some materials as the two dimensional limit is approached \cite{pracht2016,ye2012,navarro2016,saito2017}. 
We start our analysis with a brief introduction of the BdG formalism.

\section{Disordered Bogoliubov de-Gennes equations and its solutions}

The BdG equations \cite{Ghosal2001,DeGennes1964,DeGennes1966}, obtained from the saddle point solution of the path integral of a two dimensional fermionic tight binding model in a square lattice with short-range attractive interactions, are defined as follows:

\begin{equation}
\left(\begin{matrix}
	\hat{K} 		& \hat{\Delta}  \\
	\hat{\Delta}^* 	& -\hat{K}^* 	\\
\end{matrix}\right)
\left(\begin{matrix}
	u_n(r_i)   \\
	v_n(r_i) \\
\end{matrix}\right)
= E_n
\left(\begin{matrix}
	u_n(r_i)   \\
	v_n(r_i) \\
\end{matrix}\right)
\label{eq.1}
\end{equation} 

where
\begin{equation}
	\hat{K}u_n(r_i)=-t\sum_{\delta}u_n(r_i+\delta)+(V_i-\mu_i)u_n(r_i),
	\label{eq.2}
\end{equation}

$\delta$ stands for the nearest neighboring sites, $t$ is the hopping strength, $V_i$ is strength of the random potential at site $i$,  extracted from an uniform distribution  $[-V,V]$, $\mu_i = \mu + |U|n(r_i)/2$ incorporates the site-dependent Hartree shift, $\mu$ is the chemical potential and $\hat{\Delta}u_n(r_i) =\Delta(r_i)u_n(r_i)$. The same definition applies to $v_n(r_i)$. The BdG equations are completed by the self-consistency conditions for the site dependent order parameter $\Delta(r_i)$ and the density $n(r_i)$,
\begin{equation}
	\Delta(r_i) = |U|\sum_{n}u_n(r_i)v_n^*(r_i)
	\label{eq.3}
\end{equation}
and 
\begin{equation}
	n(r_i) =2\sum_{n}|v_n(r_i)|^2.
	\label{eq.4}
\end{equation}
where $U$ is the pairing interaction. 
We solve these equations for a square lattice of $N = L\times L$ sites, where $L$ is the side length of the sample and it is in units of the lattice constant. In order to minimize finite size effects, we employ periodic boundary conditions. We employ a standard iterative algorithm. Starting with an initial seed for the order parameter, we solve eq.~(\ref{eq.1}) numerically, and obtain the eigenvalues ${E_n}$ and the corresponding eigenvectors $\{u_n(r_i),v_n(r_i)\}$. We then use the self-consistent condition, eqs.~(\ref{eq.3}) and (\ref{eq.4}), to get the new value of ${\Delta(r_i)}$ and ${\mu_i}$. We repeat the process until the absolute error of ${\Delta(r_i)}$ is smaller than $10^{-6}$ or the relative error is smaller than $10^{-4}$. For convenience, all the parameters are in units of $t=1$. The numerical diagonalization was carried out in two workstations with $256$G RAM each and recent Intel Xeon multi-core processors. With this configuration, we could go to a maximum size of about $150\times 150$ sites. However, the calculation of the superfluid density is specially time consuming which prevents us to reach this limit. 

In the case of a finite Debye energy $\omega_D = 0.15$ (in units of $t$) we have found that, especially for large $V$, some disorder realizations do not converge because there are very few states inside the Debye energy window. It may be that this simply means that the only solution is the trivial one $\Delta(r_i)=0$, but we could not rule out that a non-trivial solution may exist. But it may require a very large convergence time. For this reason we only consider a disorder strength $V \leq 3$. We will see that this restriction does not impact the main results of the paper. 

\begin{figure}[htbp]
	\begin{center}
		\centering
			\includegraphics[width=12cm]{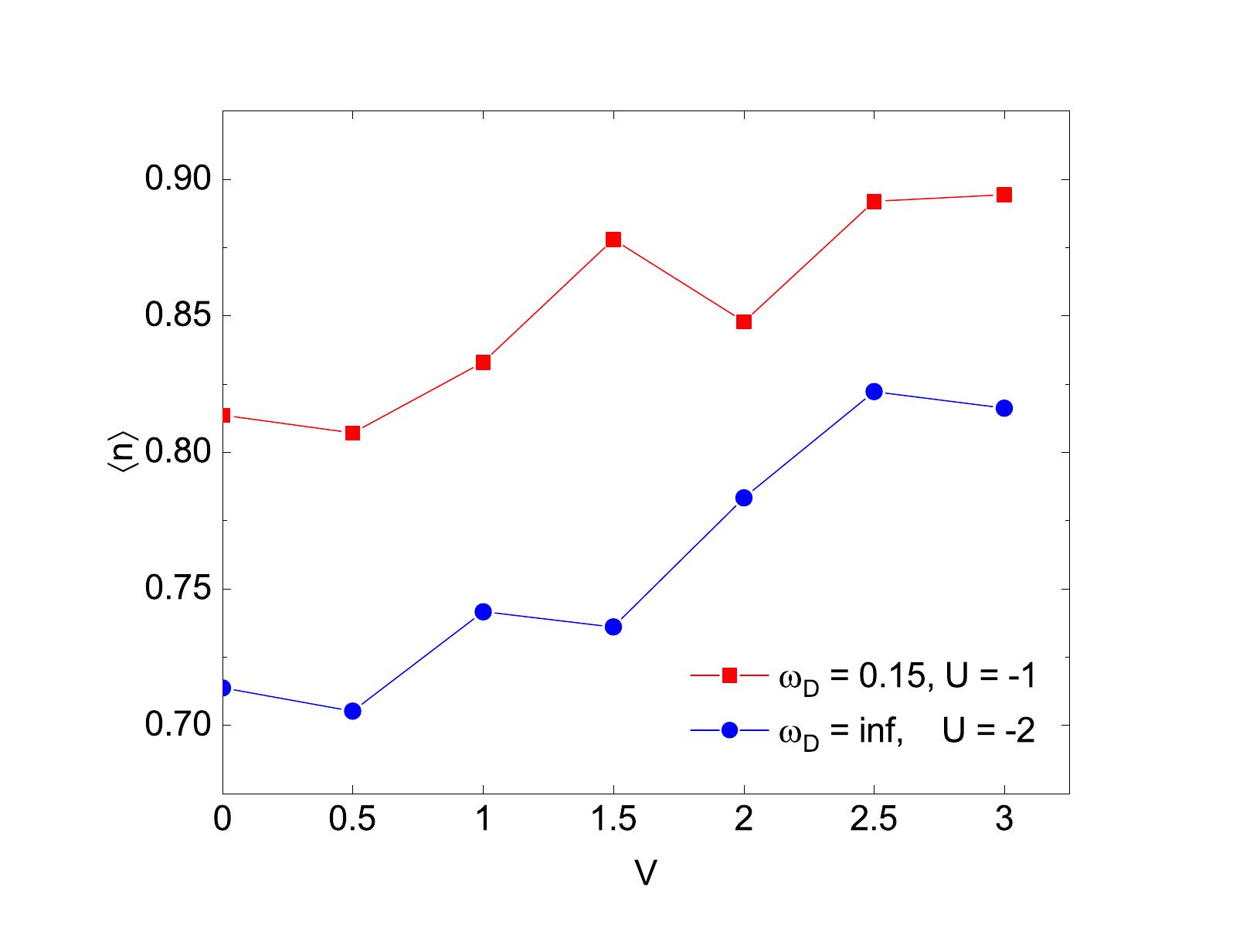}
		\caption{The electron density $\langle n \rangle$ at a fixed chemical potential for different disorder strengths. 
		As disorder increases, density fluctuations becomes larger. Fixing the chemical potential, the density is higher for stronger disorder. The system size is $80\times80$, and the chemical potential $\mu = 0$.}\label{Fig.1}
	\end{center}
\end{figure}

The averaged density $\langle n \rangle =\sum_{i}n(r_i)/N$ is determined by the chemical potential $\mu$. Here, we fixed the chemical potential at $\mu = 0$. In the strong coupling system, the states are more condensed. As is shown in Fig.~\ref{Fig.1}, if we fix $\mu$, the electron density $\langle n \rangle$ would be smaller than in the weak coupling system. By increasing disorder, electron gets more localized, which results in the increasing of $\langle n \rangle$ for a fixed $\mu$. 

\begin{figure}[htbp]
	\begin{center}
			\includegraphics[width=12cm]{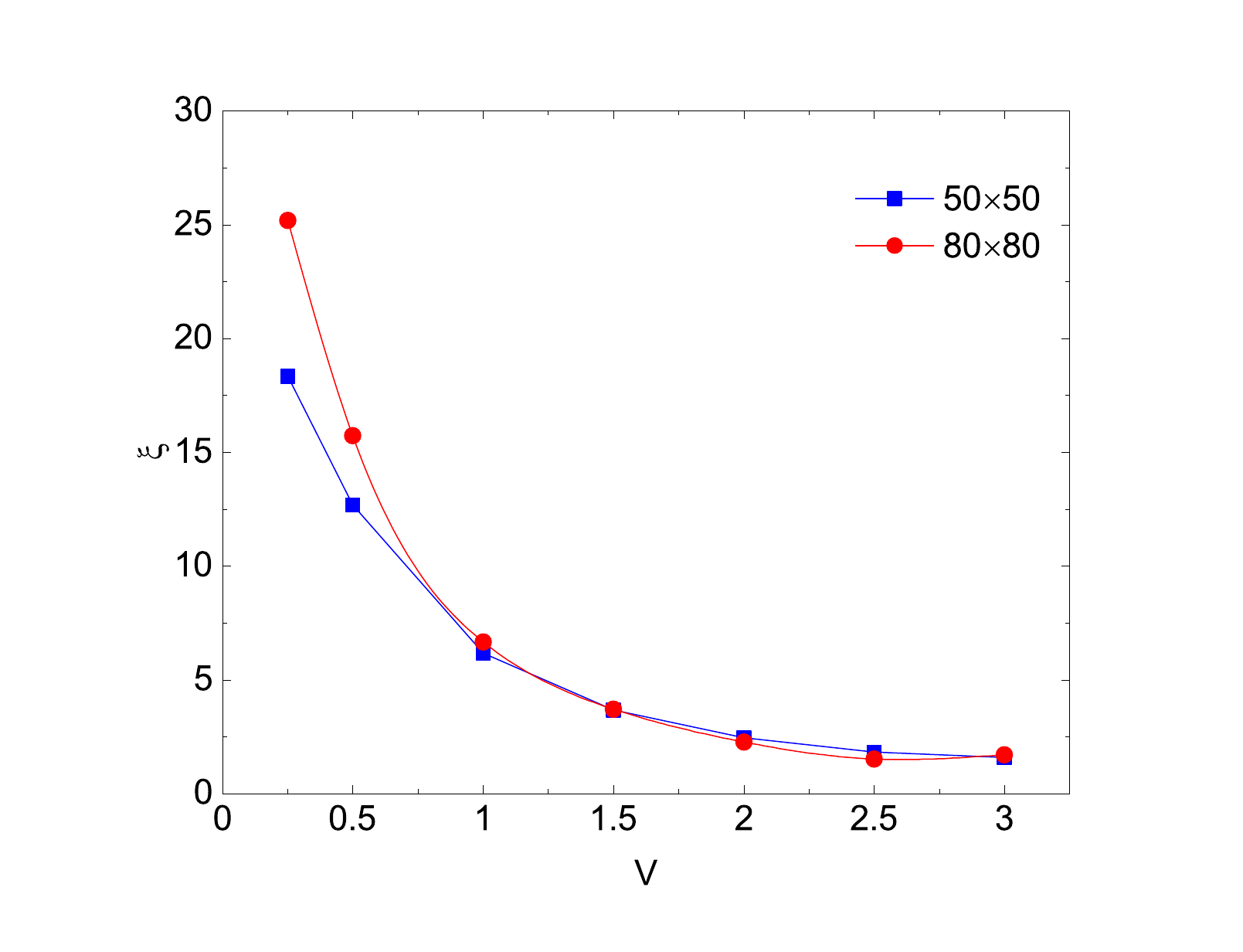}
		\caption{The superconducting coherence length $\xi$, eq.~(\ref{eq.5}), in units of the lattice spacing for different disorder strengths, two lattice sizes, Debye energy $\omega_D = 0.15$, coupling constant $U = -1$ and chemical potential $\mu = 0$, all in units of $t$. As was expected, the coherence length becomes smaller as disorder increases.  Given the maximum lattice size we can reach numerically $L \sim 150$, our results are meaningful only for $V\geq 0.5$ where $\xi \ll L$ with $L$ the typical size of the system.}\label{Fig.2}
	\end{center}
\end{figure}

One of the main motivations of the paper is to study whether the strong-coupling results of the seminal papers \cite{ghosal1998,Ghosal2001} are substantially modified for a sufficiently weak electron-phonon coupling and a finite Debye energy. This situation closer to the experimental situation of many two-dimensional superconducting materials. Obviously, for our results to be meaningful $L \gg \xi$, with  $\xi$ the superconducting coherence length, which applies for all $V$ considered. We compute $\xi$ from
\begin{equation}
\xi^2 = \frac{\int|\sum_{E_n\leq\omega_D}u_n(r)v_n(0)|^2r^2 dr}{\int |\sum_{E_n\leq \omega_D}u_n(r)v_n(0)|^2 dr} \label{eq.5}
\end{equation} 
and compare results from size $50\times50$ and size $80\times80$ in order to estimate finite size effects (see Fig.~\ref{Fig.2}). As was expected, the coherence length decrease with disorder. For $V \geq 1$, the coherence length is almost size independent and much smaller than the system size. For $80\times80$, the coherence length is still much smaller than the system size provided that $V\geq 0.5$. That limits the range of disorder we consider to $0.5 \leq V \leq 3$. 
 
\begin{figure}[htbp]
	\begin{center}
		\subfigure[]{\label{fig.s3_1} 
			\includegraphics[width=5cm]{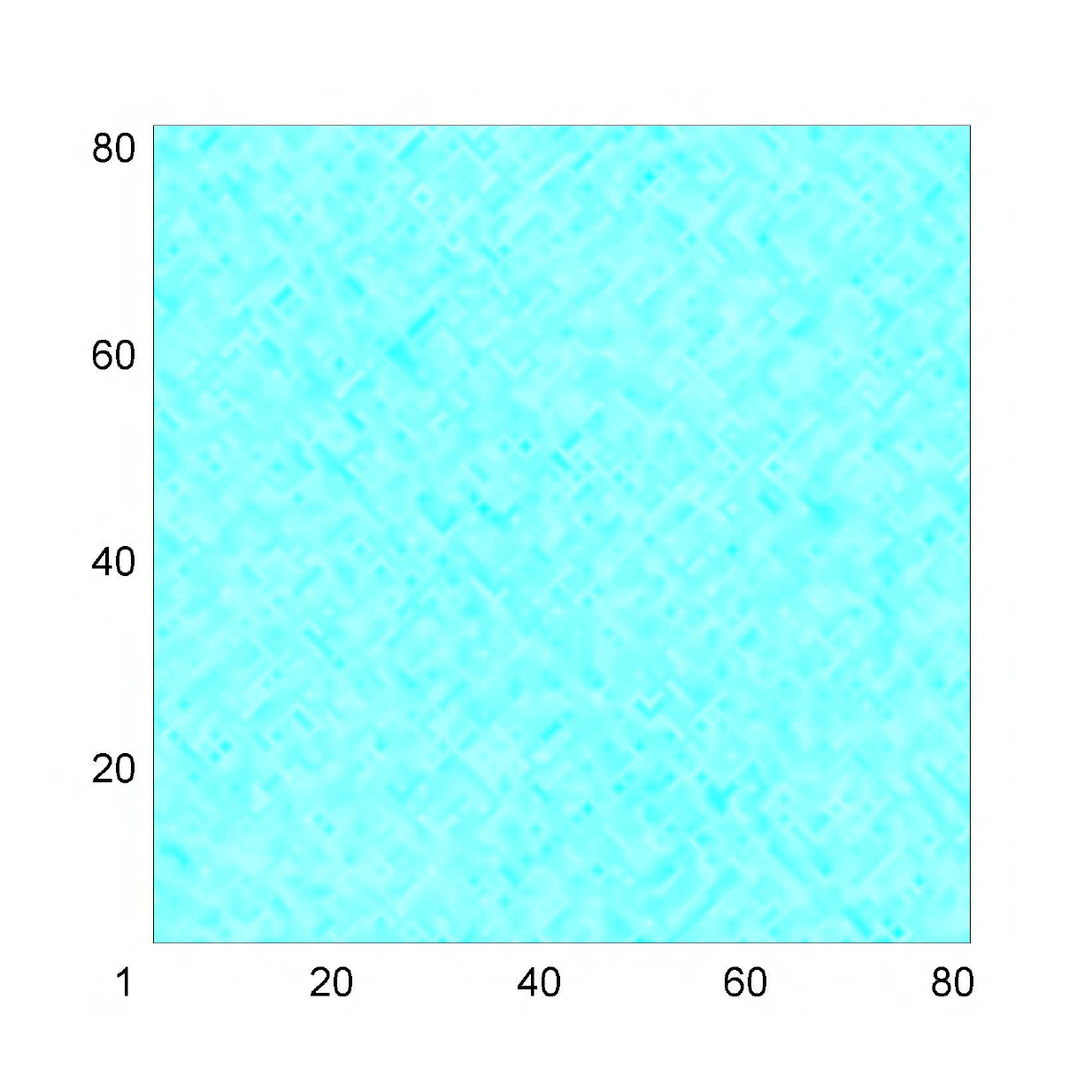}}
		\subfigure[]{\label{fig.s3_2} 
			\includegraphics[width=5cm]{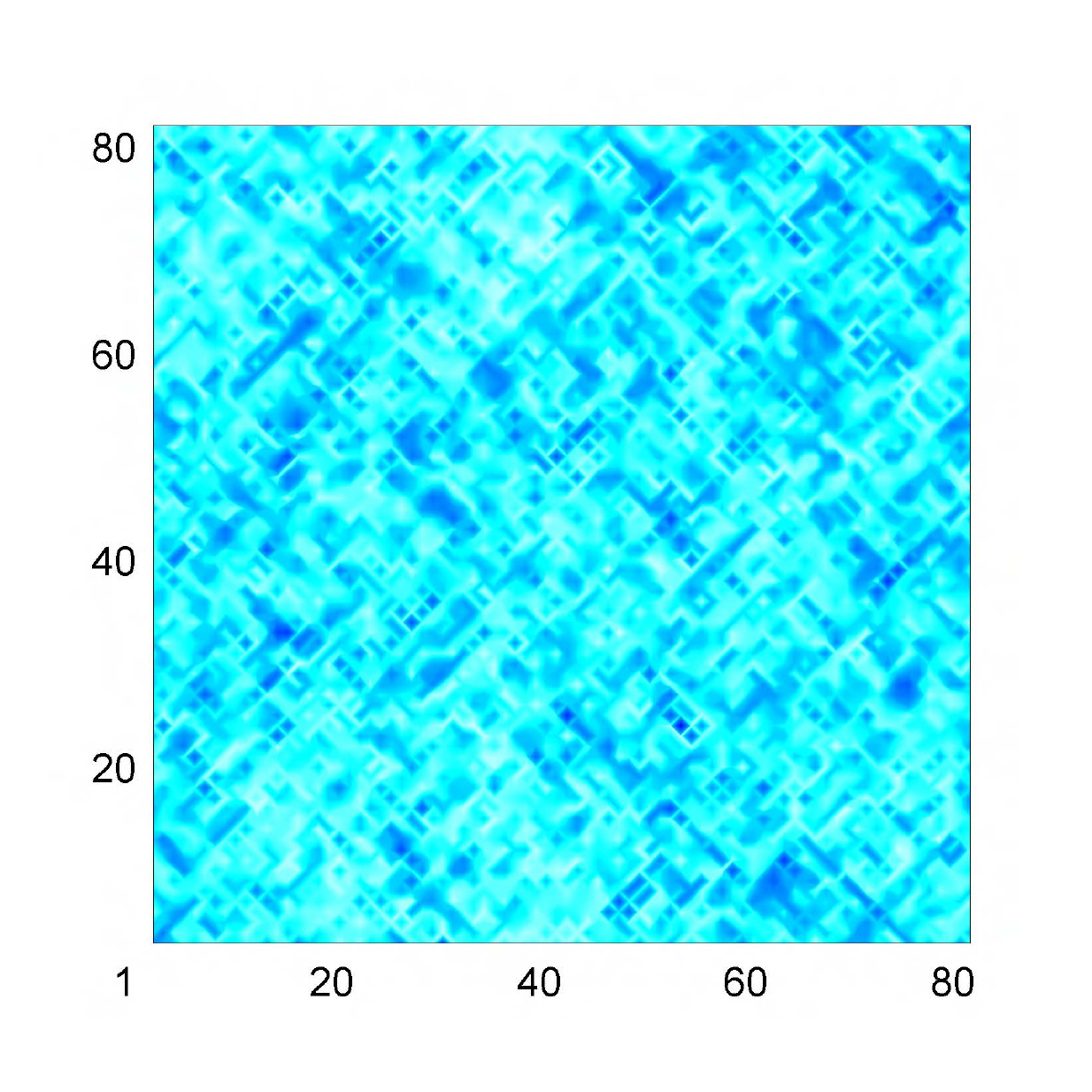}}
		\subfigure[]{\label{fig.s3_3} 
			\includegraphics[width=5cm]{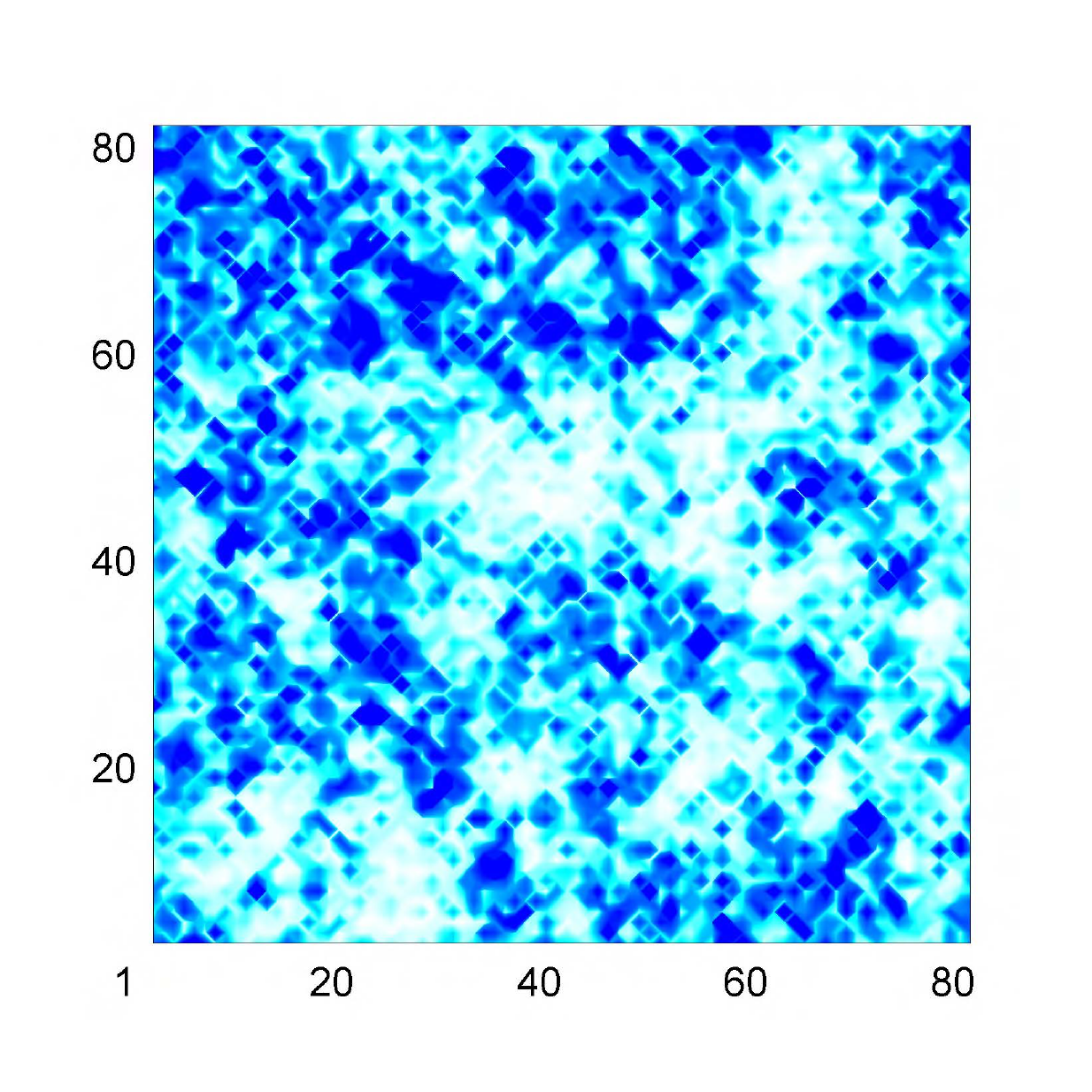}}\\
		\subfigure[]{\label{fig.s3_4} 
			\includegraphics[width=5cm]{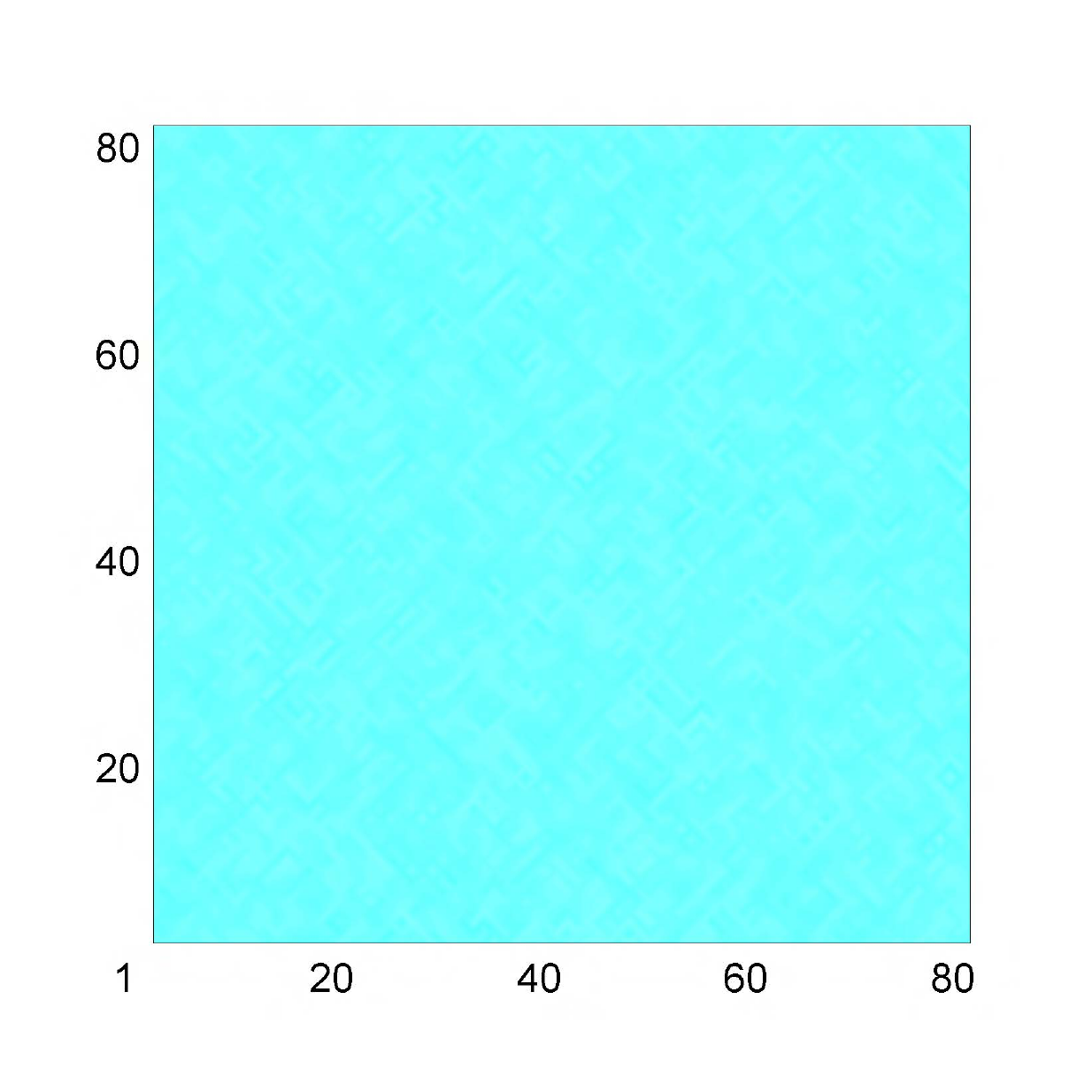}}
		\subfigure[]{\label{fig.s3_5} 
			\includegraphics[width=5cm]{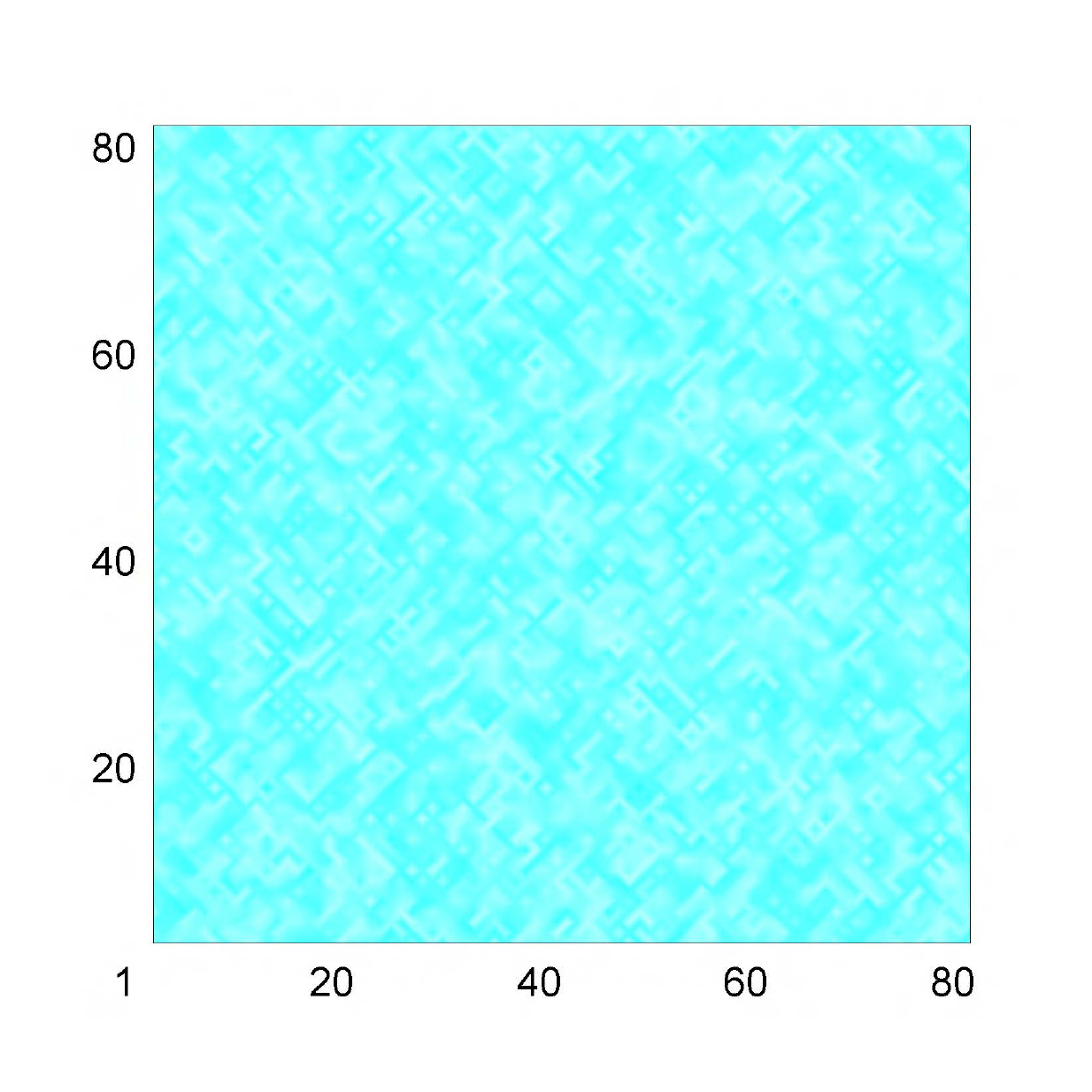}}
		\subfigure[]{\label{fig.s3_6} 
			\includegraphics[width=5cm]{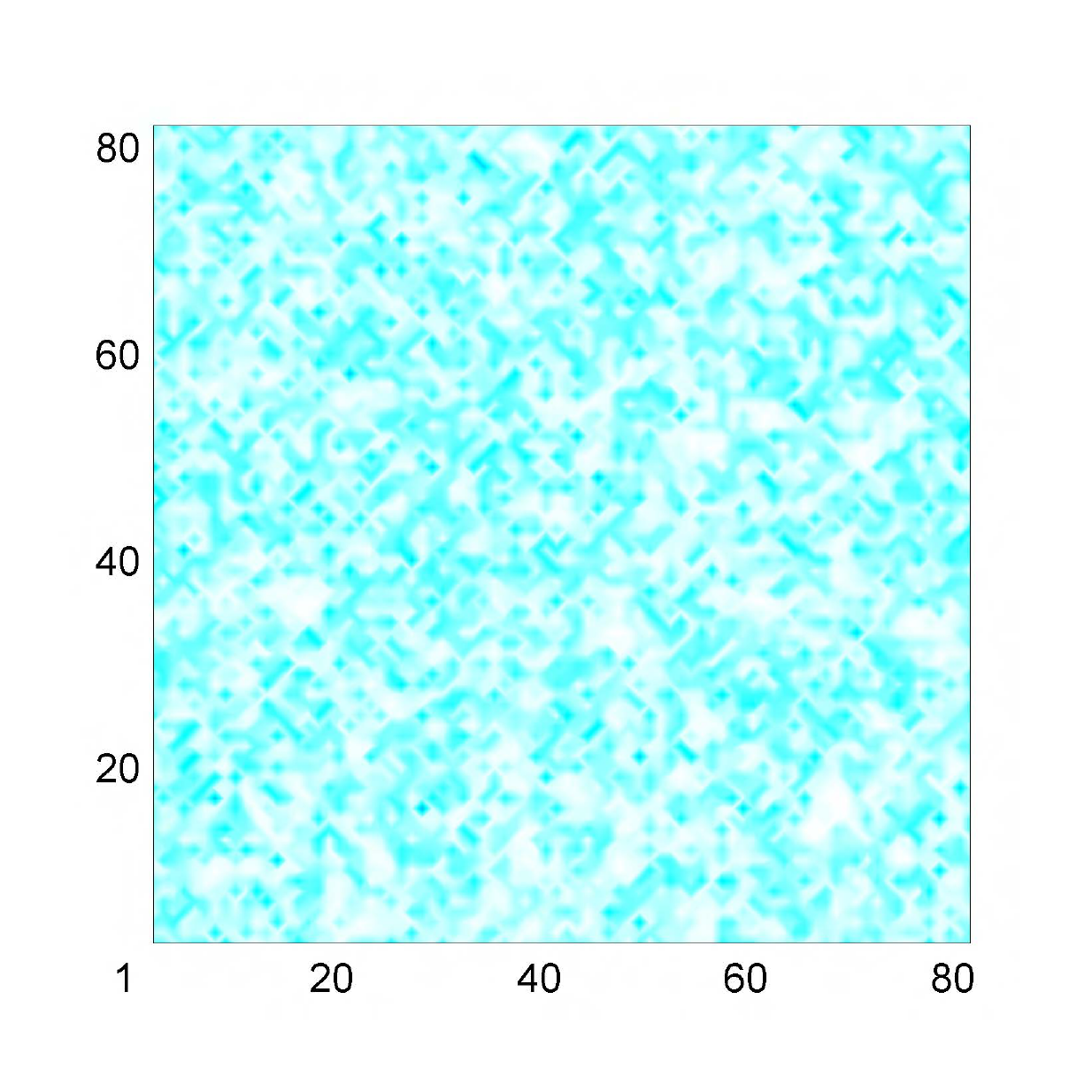}}\\
		\subfigure{}{\label{fig.s3_7} 
			\includegraphics[width=8cm]{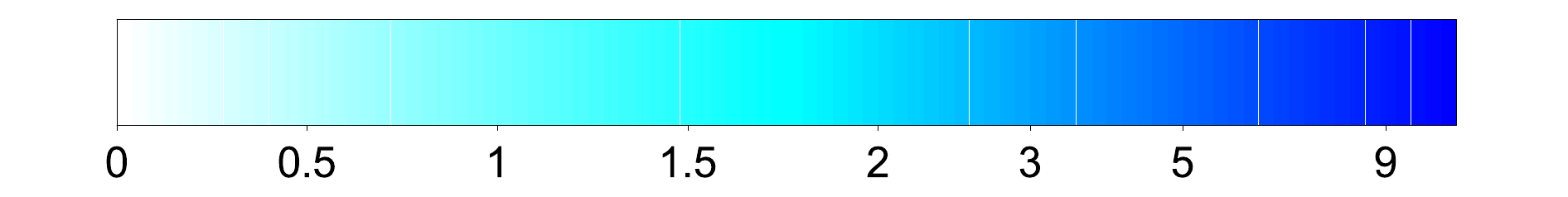}}
		\caption{The spatial distribution of the order parameter $\Delta(r_i)$ for a $80\times80$ lattice. Upper panel: Debye energy $\omega_D = 0.15$ (in units of $t$), coupling constant $U = -1$ (weak coupling).
			Lower panel: $\omega_D \to \infty$, $U = -2$ (strong coupling). The chemical potential is $\mu = 0$ in both cases. The disorder strength is $V = 0.25,1$ and $3$ from the left to right. The order parameter amplitude $\Delta(r_i)$ is normalized by $\Delta_0$, its value in the clean limit. As was expected, inhomogeneities increase with disorder though much more visibly in the weak coupling limit.}\label{Fig.3}
	\end{center}
\end{figure}

As an illustration of the type of spatial inhomogeneities induced by disorder, we show in Fig.~\ref{Fig.3} the spatial dependence of the order parameter amplitude $\Delta(r_i)$. By increasing disorder, the order parameter $\Delta(r_i)$ becomes increasingly inhomogeneous. However, there are qualitative differences between the weak coupling and the strong coupling limit. For the same disorder, the system looks less inhomogeneous in the strong coupling limit because its coherence length is smaller. Especially in Fig.~\ref{fig.s3_6}, we observe many small superconductor islands distributed in a more or less homogeneous way. By contrast, likely as a result of a much larger coherence length, spatial structure in the weak coupling region, looks much more intricate. Indeed, there are sizable spots where the order parameter is several times larger than in the no disordered limit. In the coming sections we provide more quantitative differences between these two limits. 

\section{Spatial average and probability distribution of the order parameter amplitude} 
\begin{figure}[htbp]
	\begin{center}
		\subfigure[]{\label{fig.s4_1}
			\includegraphics[width=8.5cm]{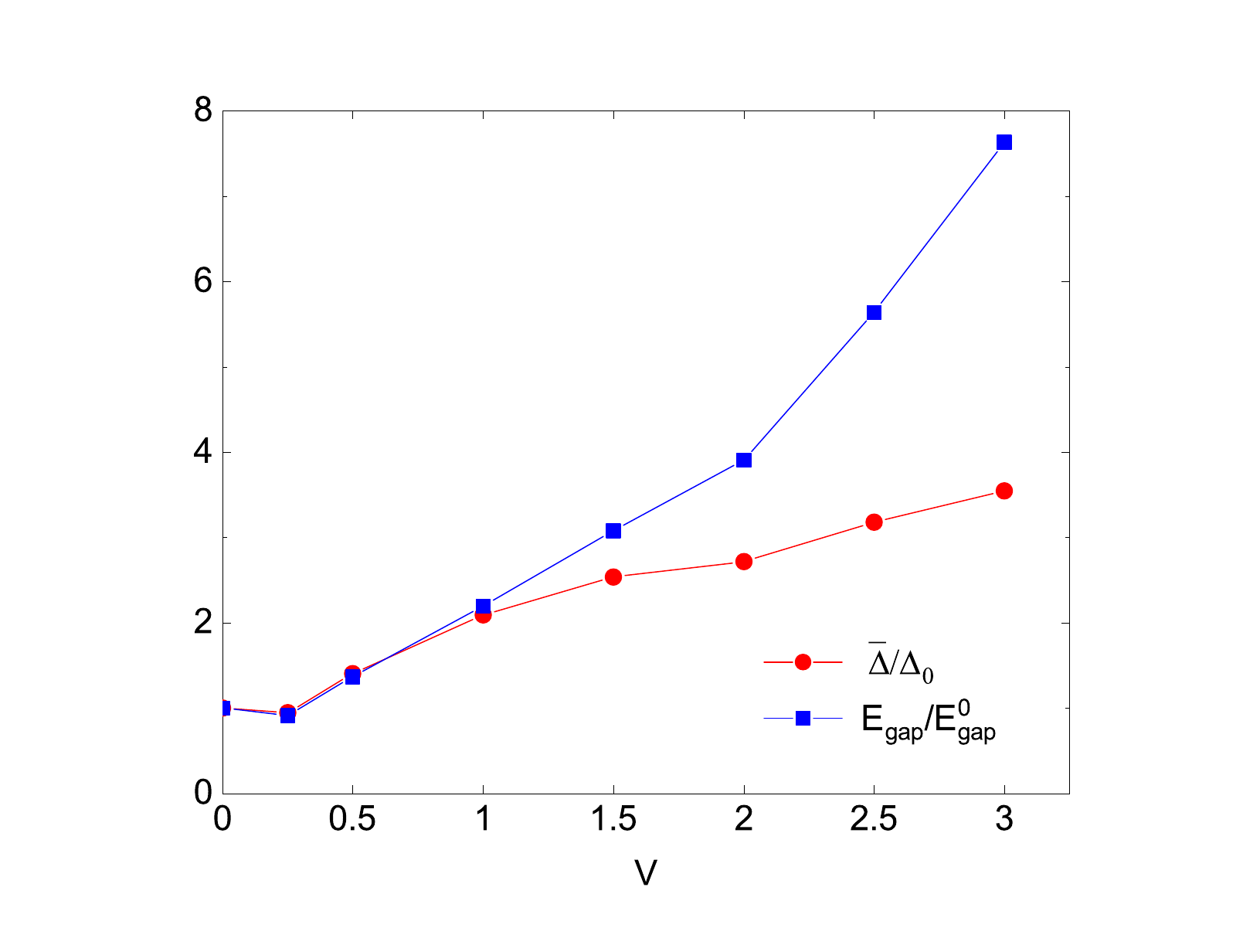}}
		\subfigure[]{\label{fig.s4_2}
			\includegraphics[width=8.5cm]{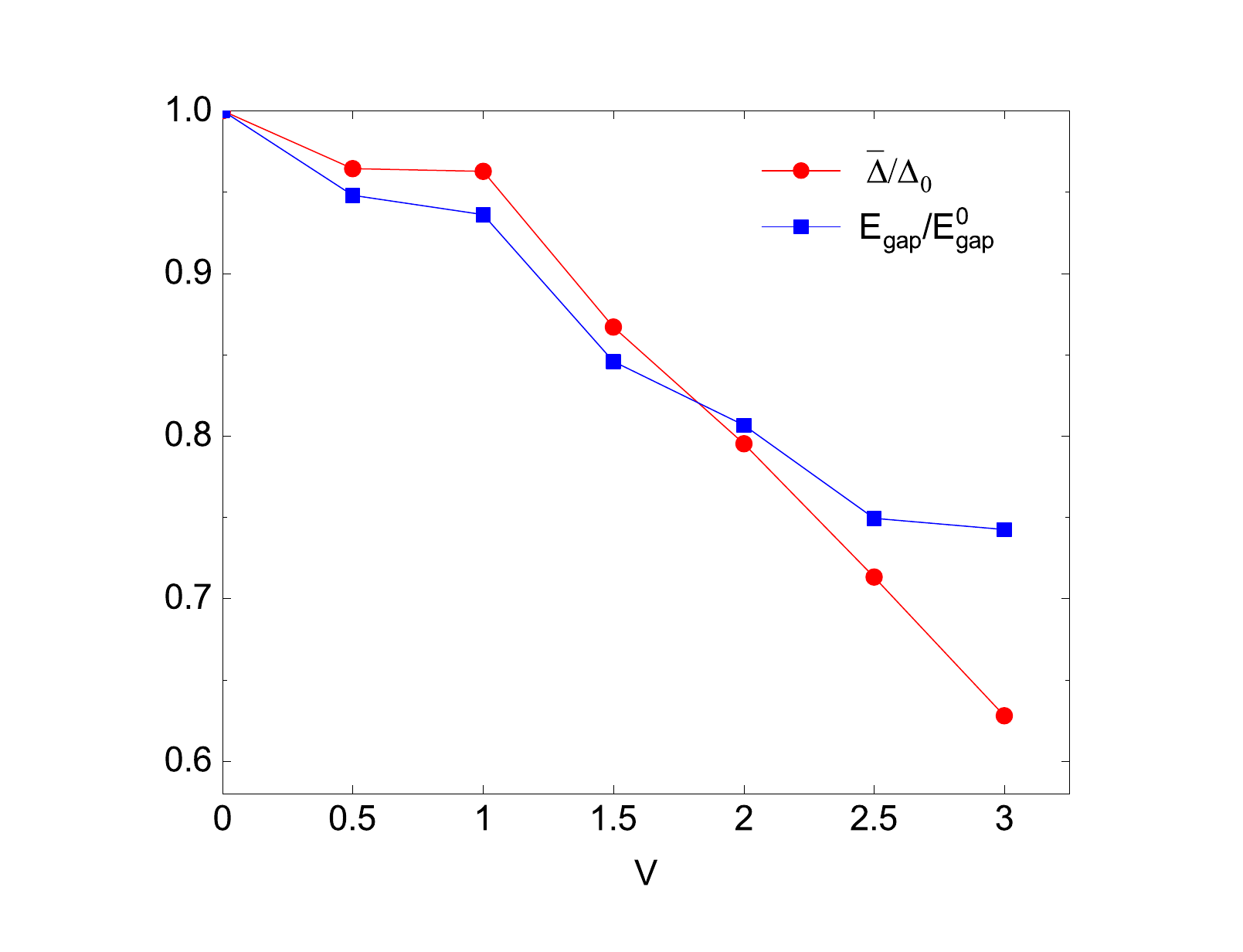}}
		\caption{The  spatial average of the order parameter $\bar{\Delta}$ and the spectral gap $E_{gap}$, obtained from the solution of the BdG equations, as a function of disorder for a size $80\times80$. \subref{fig.s4_1} weak coupling limit with Debye energy $\omega_D = 0.15$ (in units of $t$) and coupling constant $U = -1$. \subref{fig.s4_2} strong coupling limit with Debye energy $\omega_D \to \infty$ and coupling constant $U = -2$. The chemical potential is fixed at $\mu = 0$. The data is normalized by the value of the order parameter at $V = 0$. In agreement with the analytical prediction of Ref. \cite{Mayoh2015}, based on a simpler BCS approach, the average order parameter increases with disorder which suggests that disorder can enhance superconductivity.}\label{Fig.4}
	\end{center}
\end{figure}

We first compute the energy gap $E_{gap}$ and the spatial average of the order parameter amplitude $\bar{\Delta}$ which coincide in the non-disordered limit. In the strong coupling limit, it was found that \cite{Ghosal2001} the amplitude of the order parameter decreases with disorder monotonously while $E_{gap}$ decreases for weak disorder, has a minimum and then increases for sufficiently strong disorder. This increase is not related to an enhancement of superconductivity but rather to Anderson localization effects that increases the mean level spacing in an insulator. Our results, depicted in Fig.~\ref{fig.s4_2}, for larger lattices fully agree with this picture.  

However, the results for week coupling, also shown in Fig.~\ref{fig.s4_1}, are qualitatively different. $\bar{\Delta}$ increases with disorder in the range that we test numerically, but we expect that it would finally decreases for sufficiently stronger disorder. This is full agreement with the analytical prediction of Ref.~\cite{Mayoh2014a} obtained from a simpler BCS formalism, and the $T_c$ enhancement in previous numerical results\cite{Brian2018}. The energy gap $E_{gap}$ agrees with $\bar{\Delta}$ for weak disorder and then increases faster likely due to similar localization effects. These results suppose an encouraging indication that disorder may enhance superconductivity, at least for sufficiently weak disorder. The calculation later in the paper of the superfluid density will place further constraints on the conditions for an enhancement of superconductivity. However, a conclusive answer to this question would ultimately require the calculation of the critical temperature. 

\subsection{The probability distribution of the order parameter}
As was mentioned in the introduction, according to Ref.~\cite{Mayoh2015,verdu2018}, a distinctive feature of multifractal eigenstates is a log-normal distribution of the amplitude of the order parameter $\Delta(r_i)$. However, these results were obtained within a BCS formalism, which is not fully self-consistent, and only in the limit of weak-disorder and weak-coupling. Here we will show that similar findings are obtained in a fully self-consistent BdG formalism. Moreover, we compute the $f(\alpha)$ spectrum (an indicator of multifractality), related to the spatial dependence of the order parameter, to better understand to what extent the order parameter amplitude inherits the eigenstate multifractality \cite{Wegner1980,Castellani1986} observed in the non-interacting limit.

\begin{figure}[htbp]
	\begin{center}
		\centering
		\subfigure[]{\label{fig.s5_1} 
			\includegraphics[width=8.5cm]{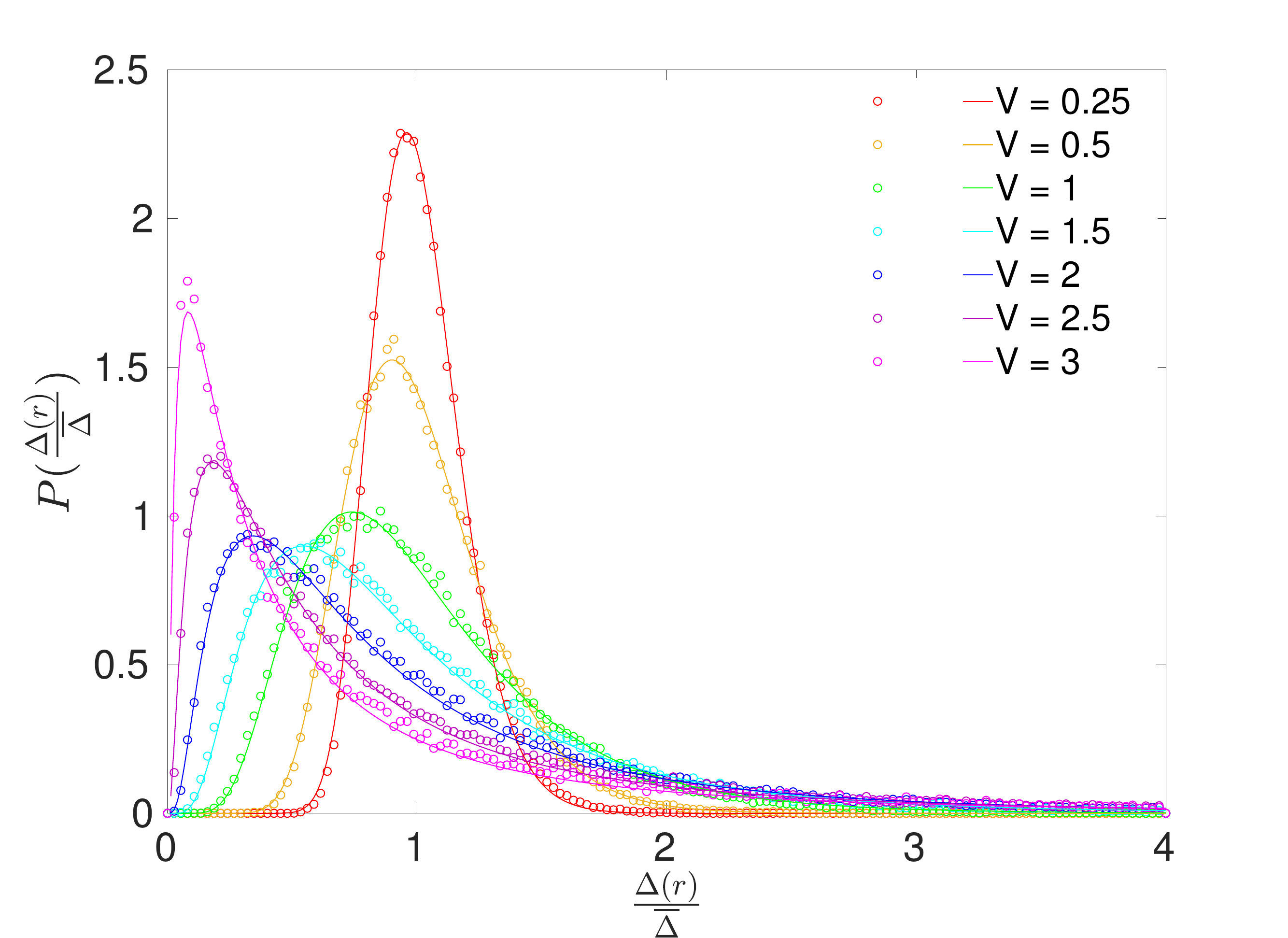}}
		\subfigure[]{\label{fig.s5_2} 
			\includegraphics[width=8.5cm]{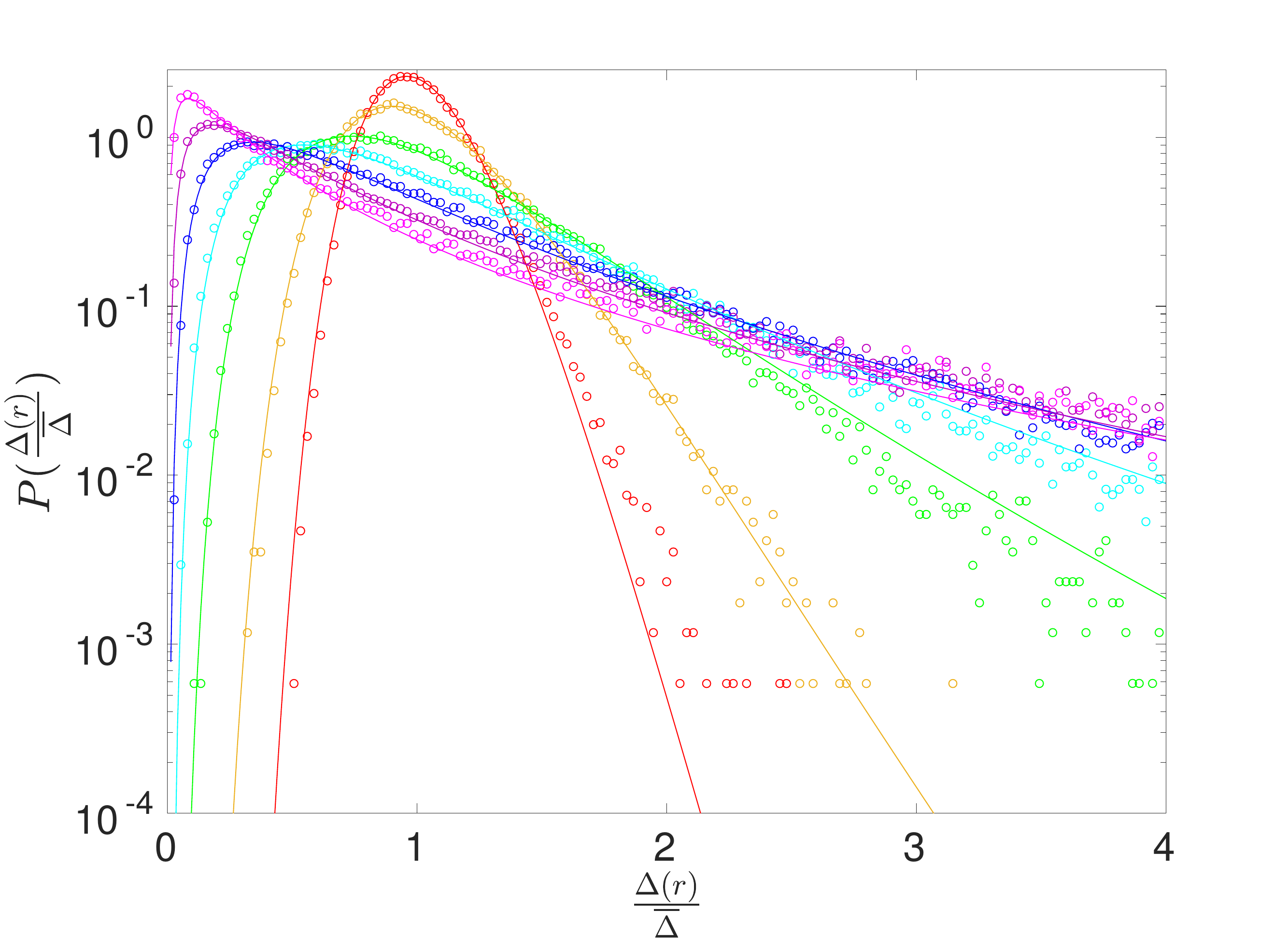}}\\
		\subfigure[]{\label{fig.s5_3} 
			\includegraphics[width=8.5cm]{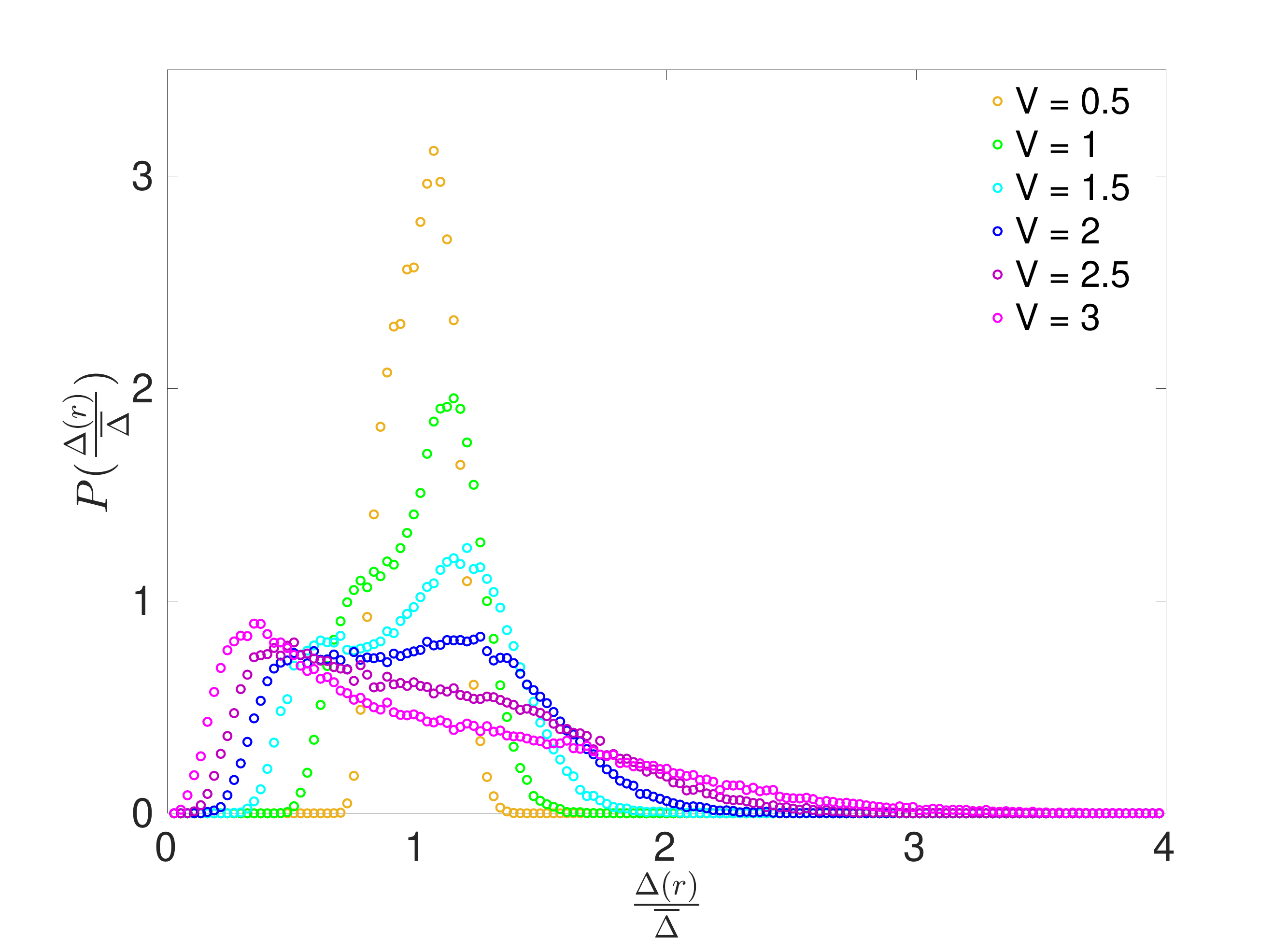}}
		\subfigure[]{\label{fig.s5_4} 
			\includegraphics[width=8.5cm]{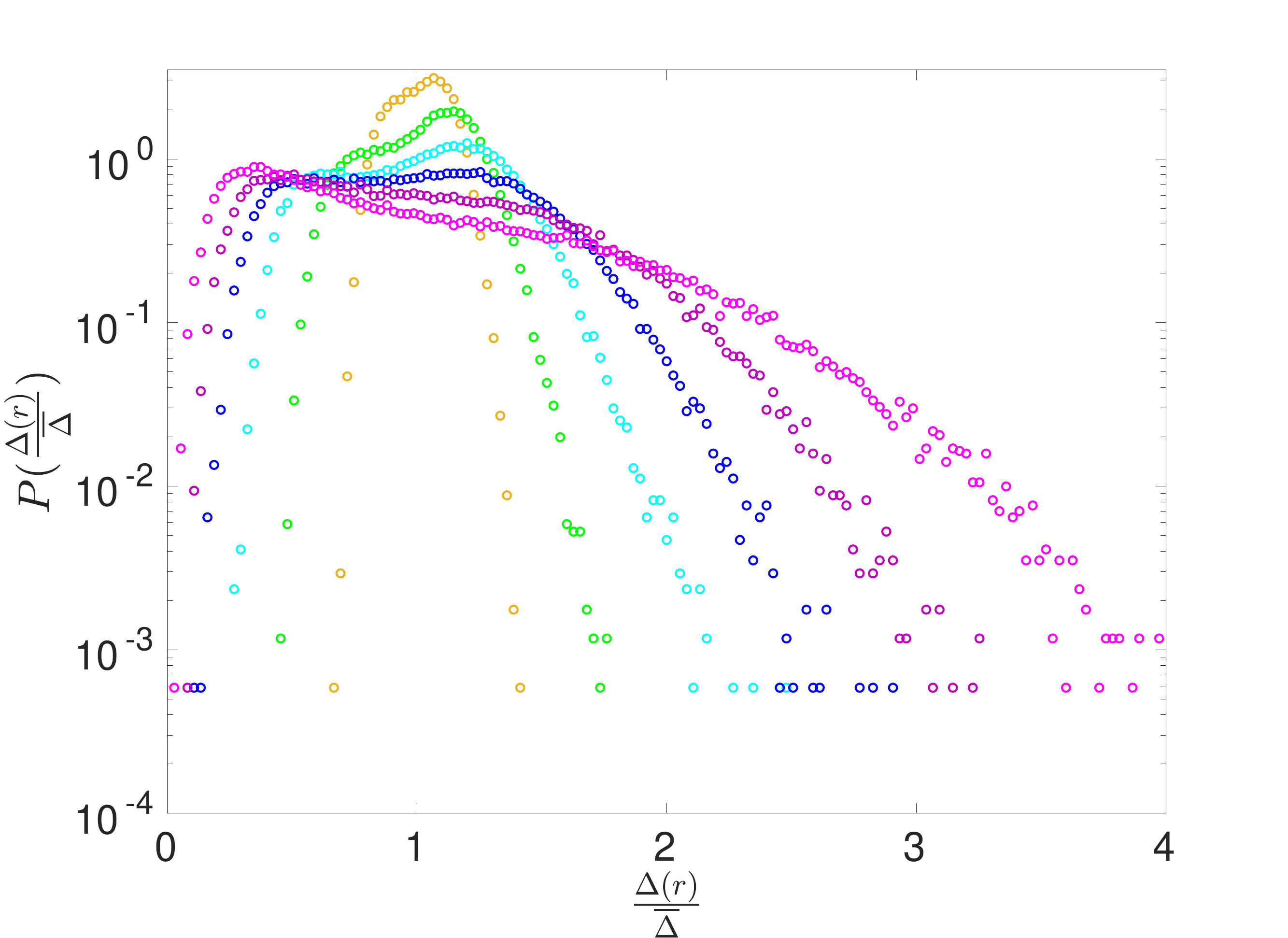}}
		\caption{The probability distribution of the order parameter $\Delta(r_i)$ normalized by its spatial average $\bar{\Delta}$. 
		The upper plots are fitted (solid lines) by the theoretical prediction \cite{Mayoh2015}, a log-normal distribution eq.~(\ref{eq.4}) that depends on two fitting parameters $\sigma$, the standard deviation, and $\mu$, the average. We find a very good agreement between the analytical expression and numerical results. The lower plots correspond to the strong coupling limit. As disorder increases, the distribution broadens as well but in a qualitatively different way. Therefore a log-normal distribution only applies in the weak-coupling limit.}\label{Fig.5}
	\end{center}
\end{figure}

The spatial probability distribution of the order parameter $P(\frac{\Delta(r)}{\bar{\Delta}})$ is plotted in Fig.~\ref{Fig.5} as function of disorder. In the BCS limit, and assuming that the eigenfunction correlation in the non-interacting limit are multifractal, it was found \cite{Mayoh2015} that the probability distribution of the order parameter in the weak coupling limit is log-normal,
\begin{equation}
P\left(\frac{\Delta(r)}{\bar{\Delta}}\right) = \frac{\bar{\Delta}}{\Delta(r)\sqrt{2\pi}\sigma}\exp\left(-\frac{\left[\lg\left(\frac{\Delta(r)}{\bar{\Delta}}\right)-\mu\right]^2}{2\sigma^2}\right).	\label{eq.6}
\end{equation}

In the weak coupling limit, we observe a very good agreement with this log-normal distribution, using the standard deviation $\sigma$ and average $\mu$ as fitting parameters, and the numerical results. When disorder is small, the distribution is narrow and centered on the mean gap $\bar{\Delta}$. As the strength of disorder increases, the distribution of $\frac{\Delta(r)}{\bar{\Delta}}$ becomes increasingly broader and asymmetric. 
Interestingly, in the strong coupling limit, the results are qualitatively different. A log-normal distribution is never a good fit of the data which suggests that multifractality only plays a role in the weak coupling limit. 
Finally, we note that in the strong disorder limit the probability distribution is closer to Poisson distribution for both weak and strong coupling. We shall show in this limit phase coherence is lost but the system is still in the metallic phase. 

\subsection{Singularity spectrum of the order parameter amplitude}
In order to have a more quantitative understanding of the role of multifractality in our system, we compute the $f(\alpha)$ spectrum (see \cite{PhysRevLett.62.1327} for definition and details about its calculation), also called singularity spectrum, of $\Delta(r_i)$ that provides information about its range of scaling exponents. In the limit of weak disorder, it was found that the $f(\alpha)$ spectrum of the density of probability of isolated eigenstates of the 2d Anderson model, for sizes much smaller than the localization length, was parabolic which is a feature of a multifractal measure. Indeed, $\Delta(r_i)$ is given by the self-consistent condition, eqs.~(\ref{eq.3}) and (\ref{eq.4}), which is a weighted average over the eigenstates $u_n(r_i)$ and $v_n(r_i)$ of the BdG equations. At least for clean nano-grains \cite{Shanenko2007}, it was found that $u_n(r_i)$ and $v_n(r_i)$  are proportional to the eigenstates of the one-body problem $\Psi_n(r_i)$ for sufficiently weak coupling. Therefore, it seems plausible, especially if the weighted sum defining $\Delta(r_i)$ does not contain many eigenstates, that some of the anomalous scaling features of the eigenstates of the one-body problem may also be present in the order parameter. More specifically, we define $|P(r_i)|^2 = \frac{\Delta(r_i)}{\sum_{j=1}\Delta(r_j)}$ and compute the $f(\alpha)$ spectrum of $|P(r_i)|^2$ following the method introduced in Ref.~\cite{PhysRevLett.62.1327}. 

\begin{figure}[htbp]
	\begin{center}
		\centering
		\subfigure[]{\label{fig.s6_1} 
			\includegraphics[width=5cm]{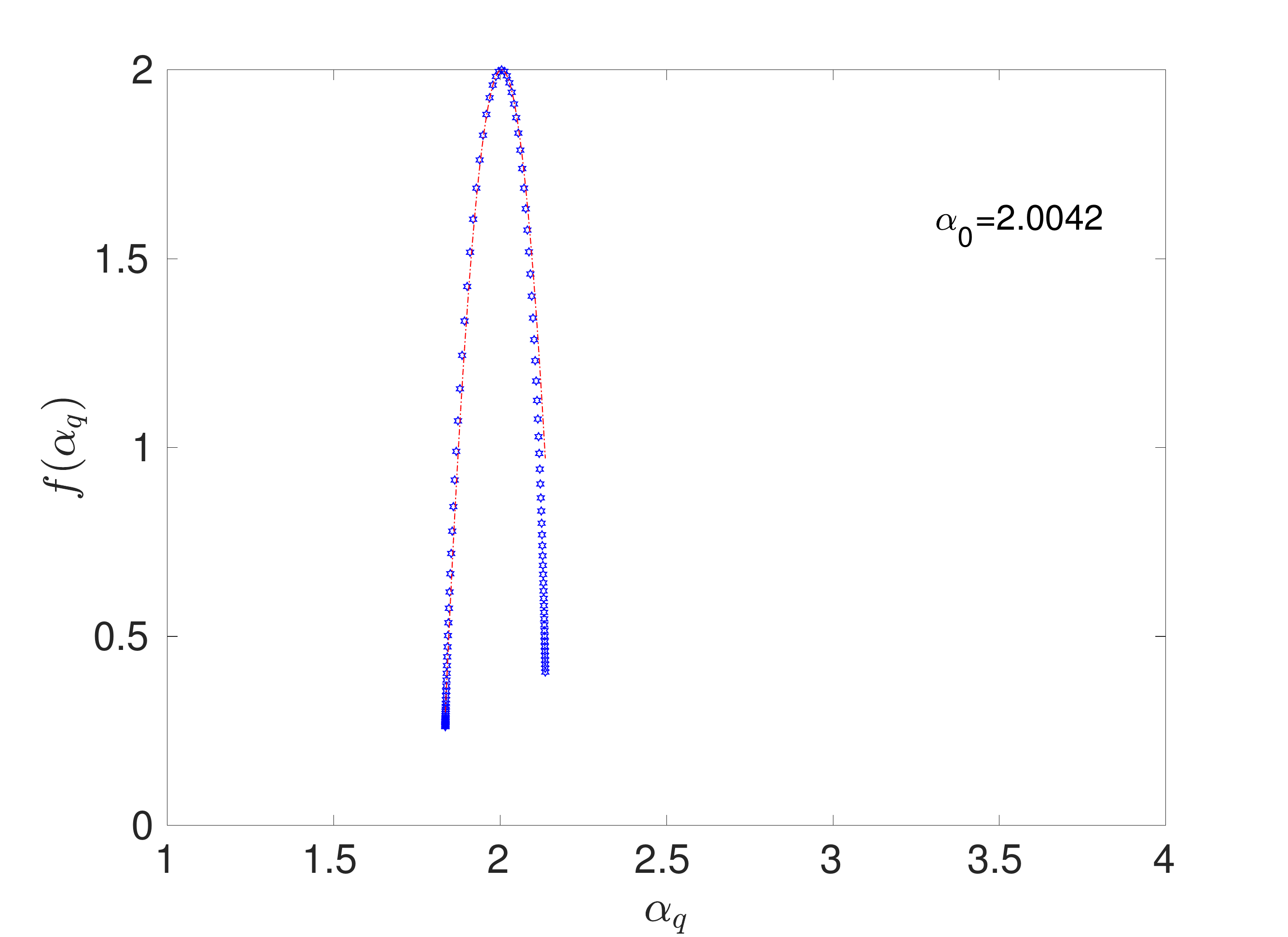}}
		\subfigure[]{\label{fig.s6_2} 
			\includegraphics[width=5cm]{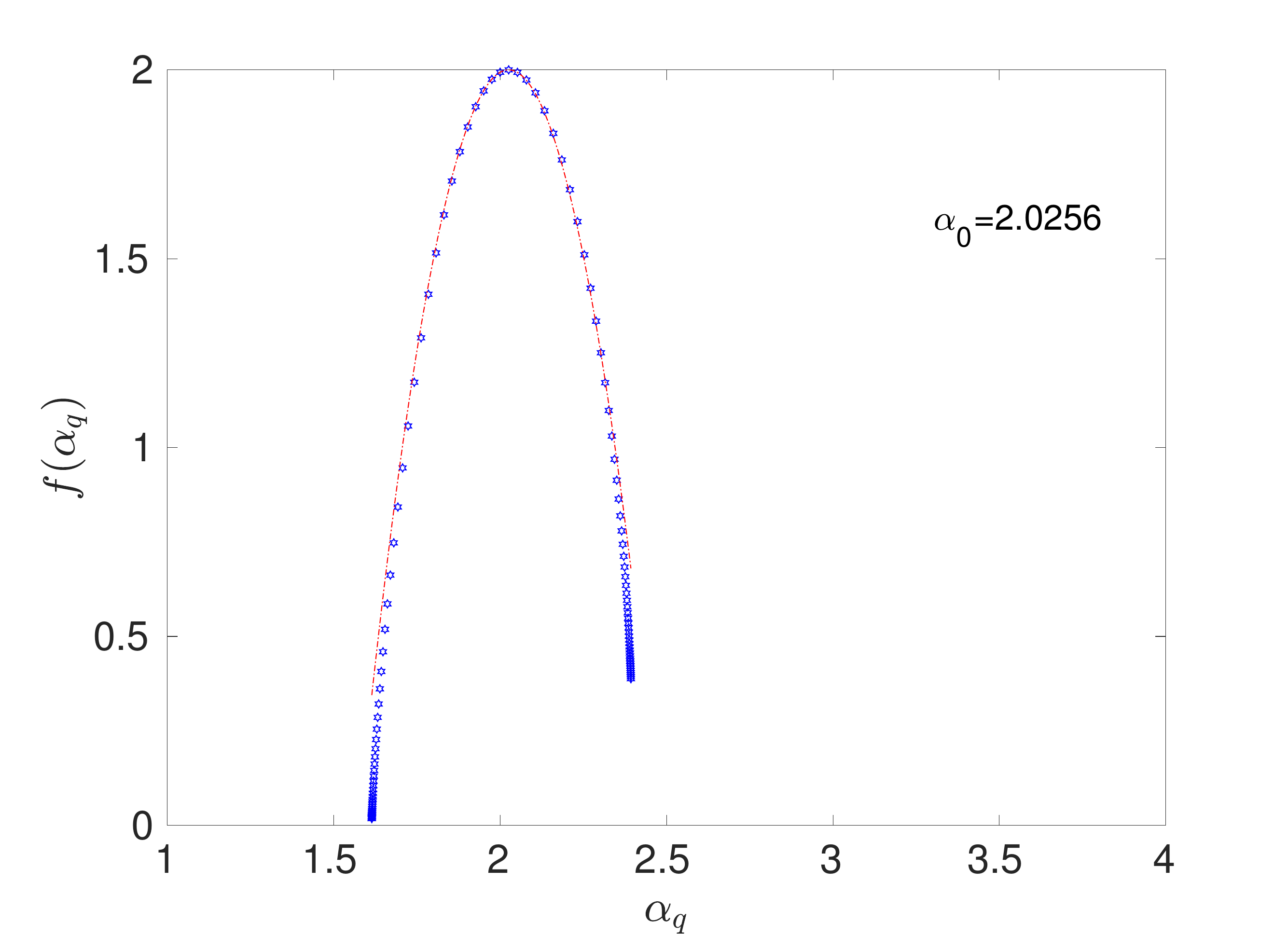}}
		\subfigure[]{\label{fig.s6_3} 
			\includegraphics[width=5cm]{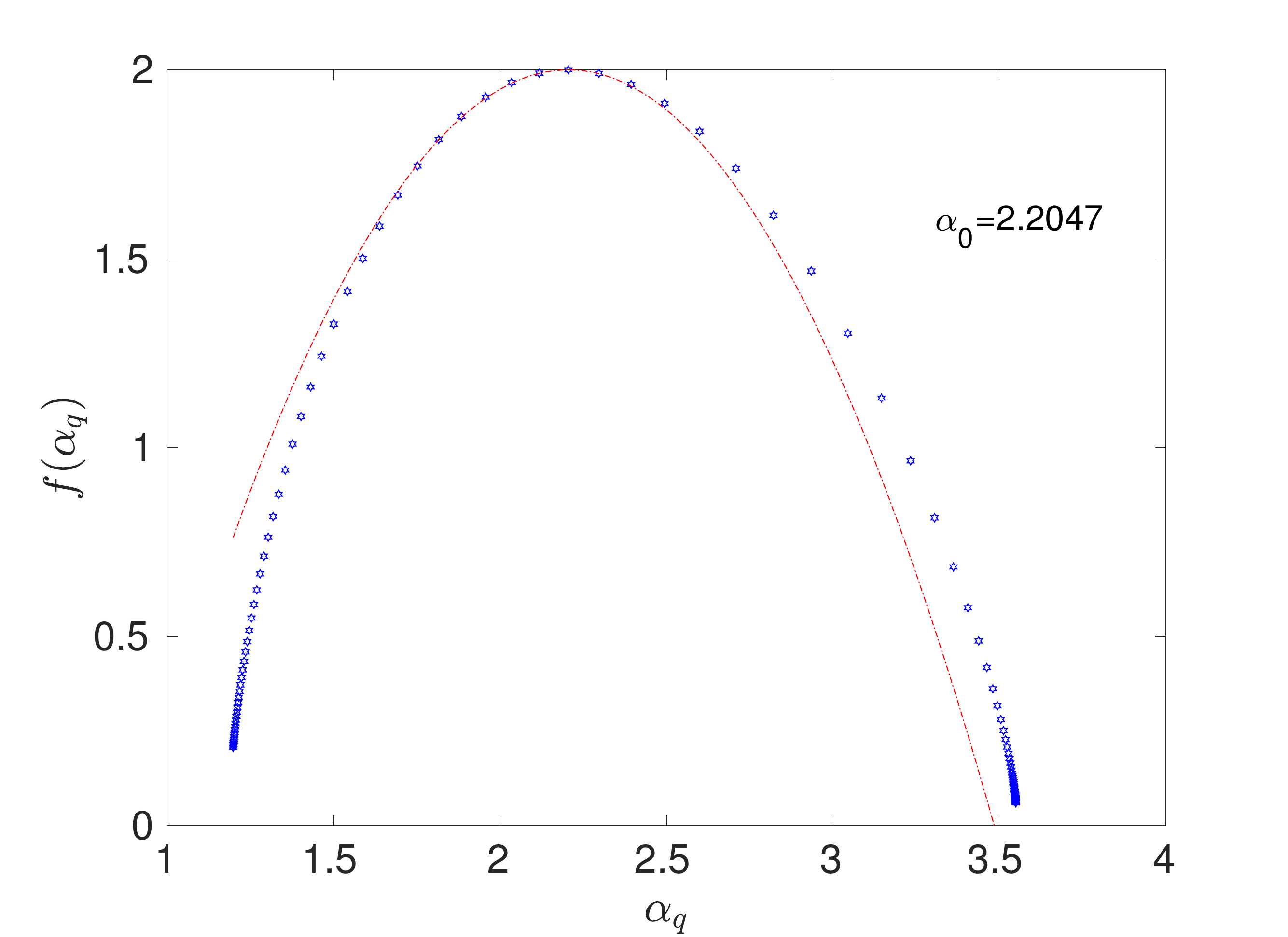}}\\
		\subfigure[]{\label{fig.s6_4} 
			\includegraphics[width=5cm]{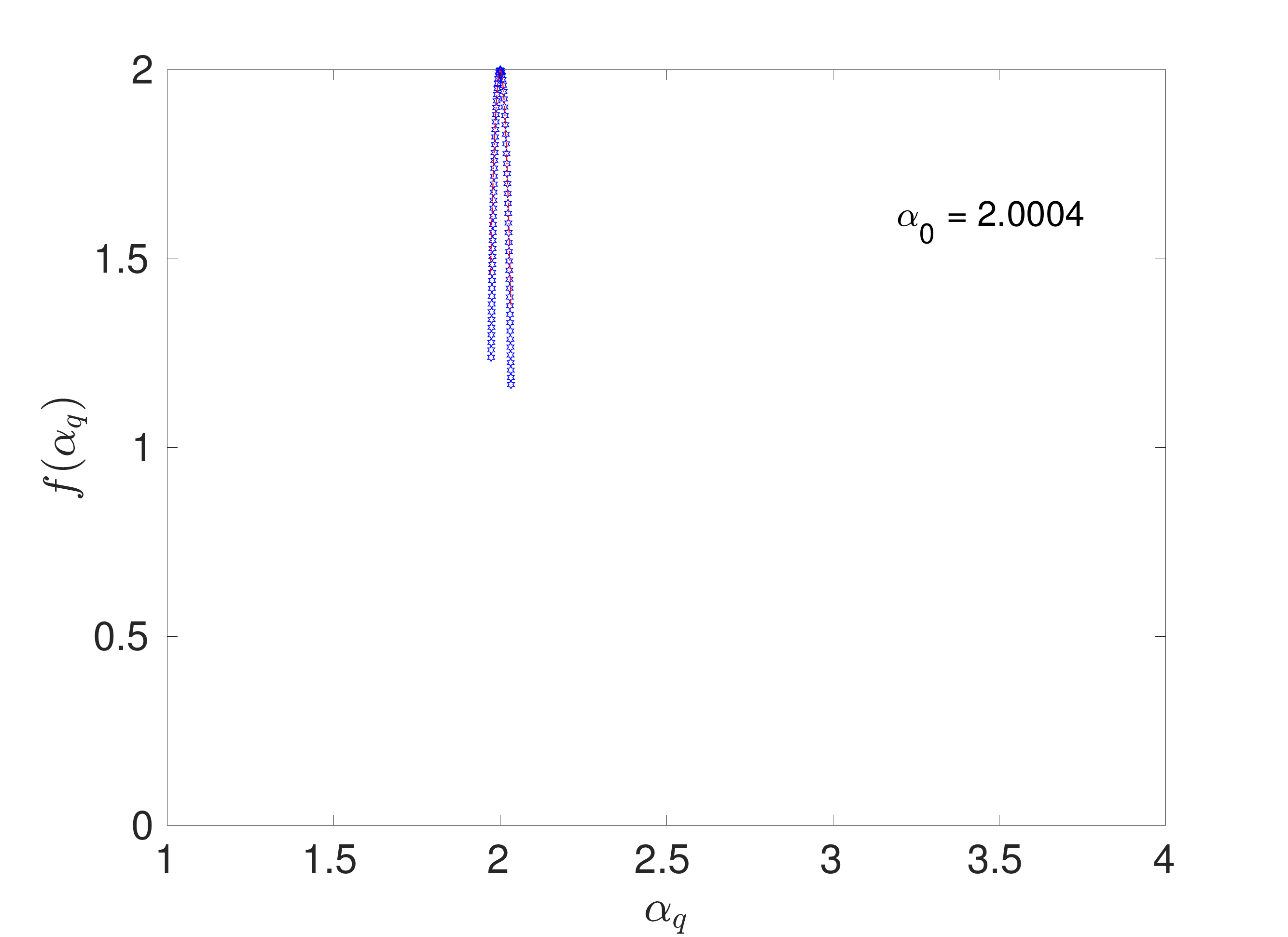}}
		\subfigure[]{\label{fig.s6_5} 
			\includegraphics[width=5cm]{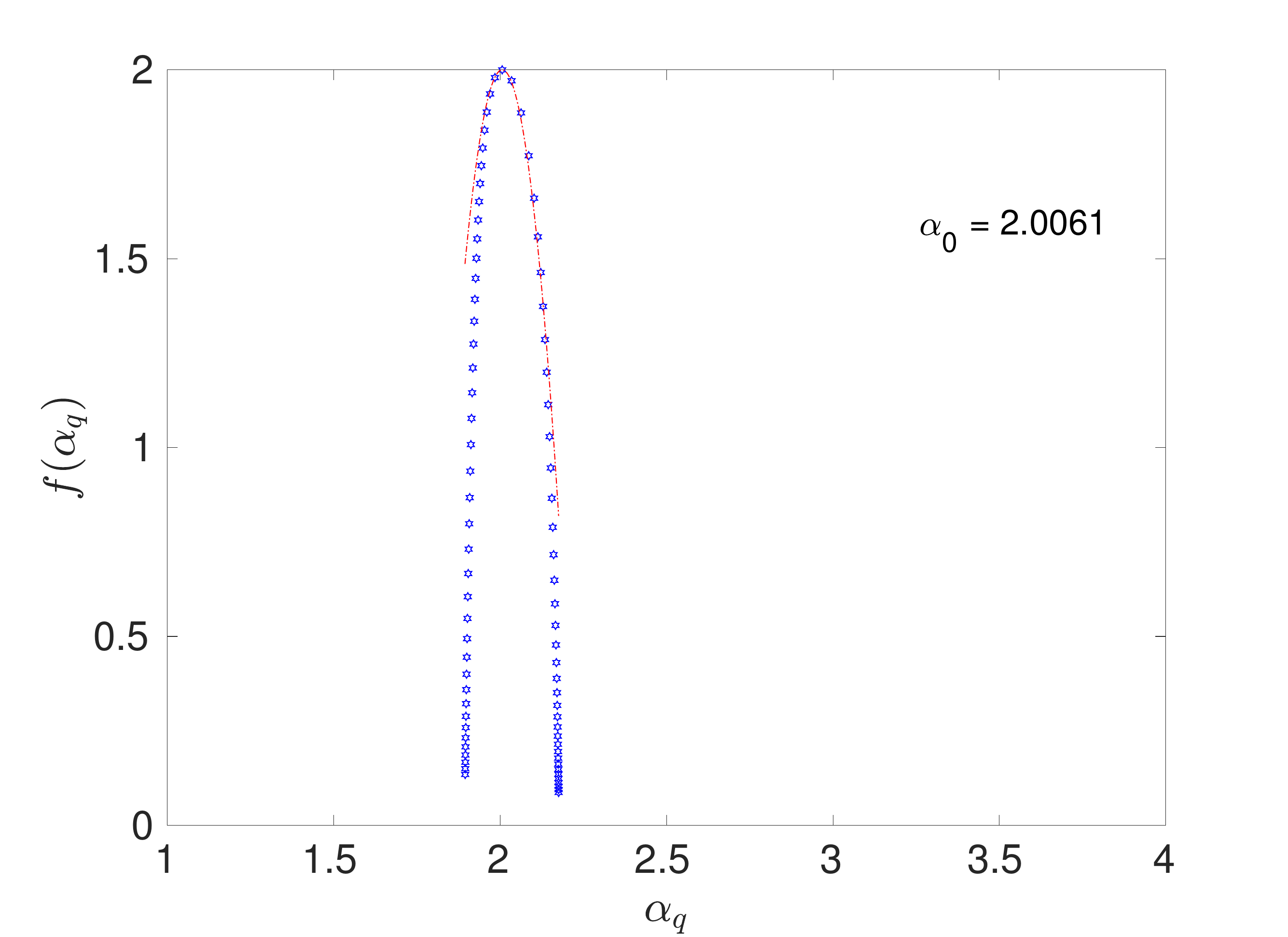}}
		\subfigure[]{\label{fig.s6_6} 
			\includegraphics[width=5cm]{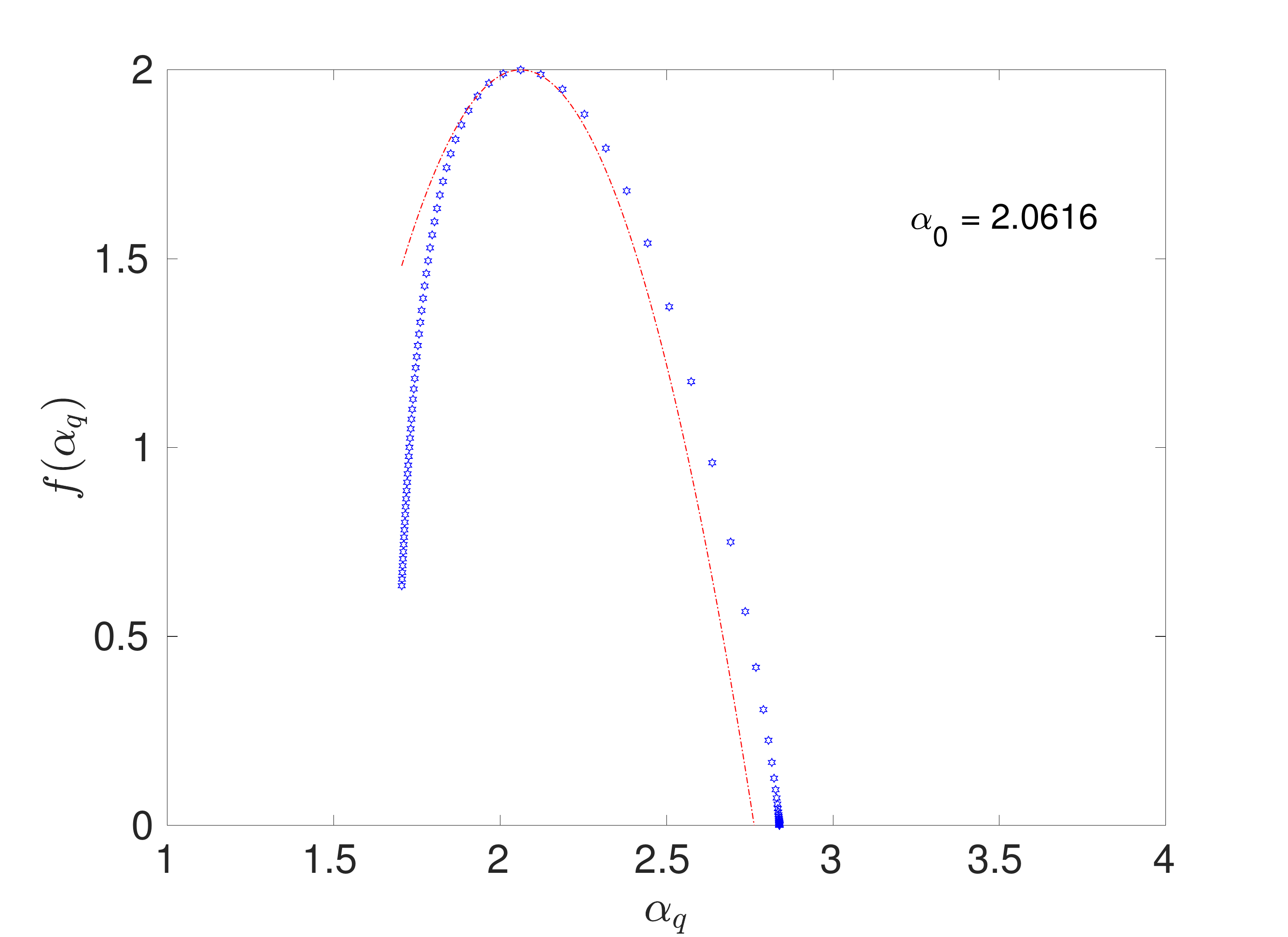}}
		\caption{Singularity spectrum $f(\alpha)$ related to the order parameter $\Delta(r_i)$ for the parameters of Fig.~\ref{Fig.3}. 
		In the weak coupling limit (upper panel), it agrees well with the parabolic prediction (solid line) corresponding to weakly multifractal eigenstates. Also in agreement with the theoretical prediction, the parabolic curve becomes broader and its maximum shifts to larger values as disorder increases.  In the strong coupling limit (lower panel), multifractal effects are much harder to observe. The distribution is rather narrow and the agreement with a parabola is worse which suggests that, at least in the studied disorder range, the singularity spectrum has no multifractal-like features.}\label{Fig.6}
	\end{center}
\end{figure}

In Fig.~\ref{Fig.6}, we present results for $f(\alpha)$, for the parameters employed in Fig.~\ref{Fig.3}. We find that in the weak-disorder, weak-coupling region, the singularity spectrum $f(\alpha)$ is well approximated by  $f(\alpha) = 2 - \frac{(\alpha - \alpha_0)^2}{4(\alpha_0 - 2)}$, with $\alpha_0 =2 +1/g$ and $g$ the dimensionless conductance. This is the analytical prediction for the density of probability related to weakly multifractal eigenstates \cite{Evers2008} in two dimensions. In our case $\alpha_0$ is a fitting parameter. As disorder increases, $\alpha_0$ increases as well, and the parabolic curve becomes broader, which means the spatial distribution of the order parameter is increasingly inhomogeneous. 
This is another indication that the spatial distribution of order parameter has an intricate spatial structure reminiscent of a multifractal measure.

The situation is different in the strong coupling limit. The fitting to a parabola is in general worse. Moreover, the fitted values of $\alpha_0$ are very close to $2$ and the $f(\alpha)$ spectrum is much narrower than in the weak-coupling region. This in an indication of a much more homogeneous distribution which is not really multifractal. In summary, multifractality is a feature only attached to weakly-coupled, weakly-disordered superconductors. We now investigate in more detail this issue by looking directly at the eigenvectors $u_n(r_i)$ and $v_n(r_i)$ in order to determine how the order parameter is constructed from them.

\subsection{Overlap between $u_n(r)$ and $v_n(r)$}

In order to gain a more quantitative understanding on the relation between $\Delta(r)$ and eigenfunctions of the BdG equation $\{u_n(r),v_n(r)\}$, we study
\begin{equation} \label{puv} P_{uv} = \sum_{r}|u_n^2(r)-v_n^2(r)| \end{equation}
When $u_n$ and $v_n$ overlap strongly, then $P_{uv} \approx 0$ while if $u_n$ and $v_n$ are completely decoupled then $P_{uv} \approx 1$ since $\sum_{r}(u_n^2(r)+v_n^2(r)) = 1$. We note that the self-consistent condition eq.~(\ref{eq.3}) dictates that only eigenstates $u_n$ and $v_n$ that overlap strongly contribute substantially to $\Delta(r)$. 

\begin{figure}[htbp]
	\begin{center}
		\subfigure[]{\label{fig.s7_1} 
			\includegraphics[width=4cm]{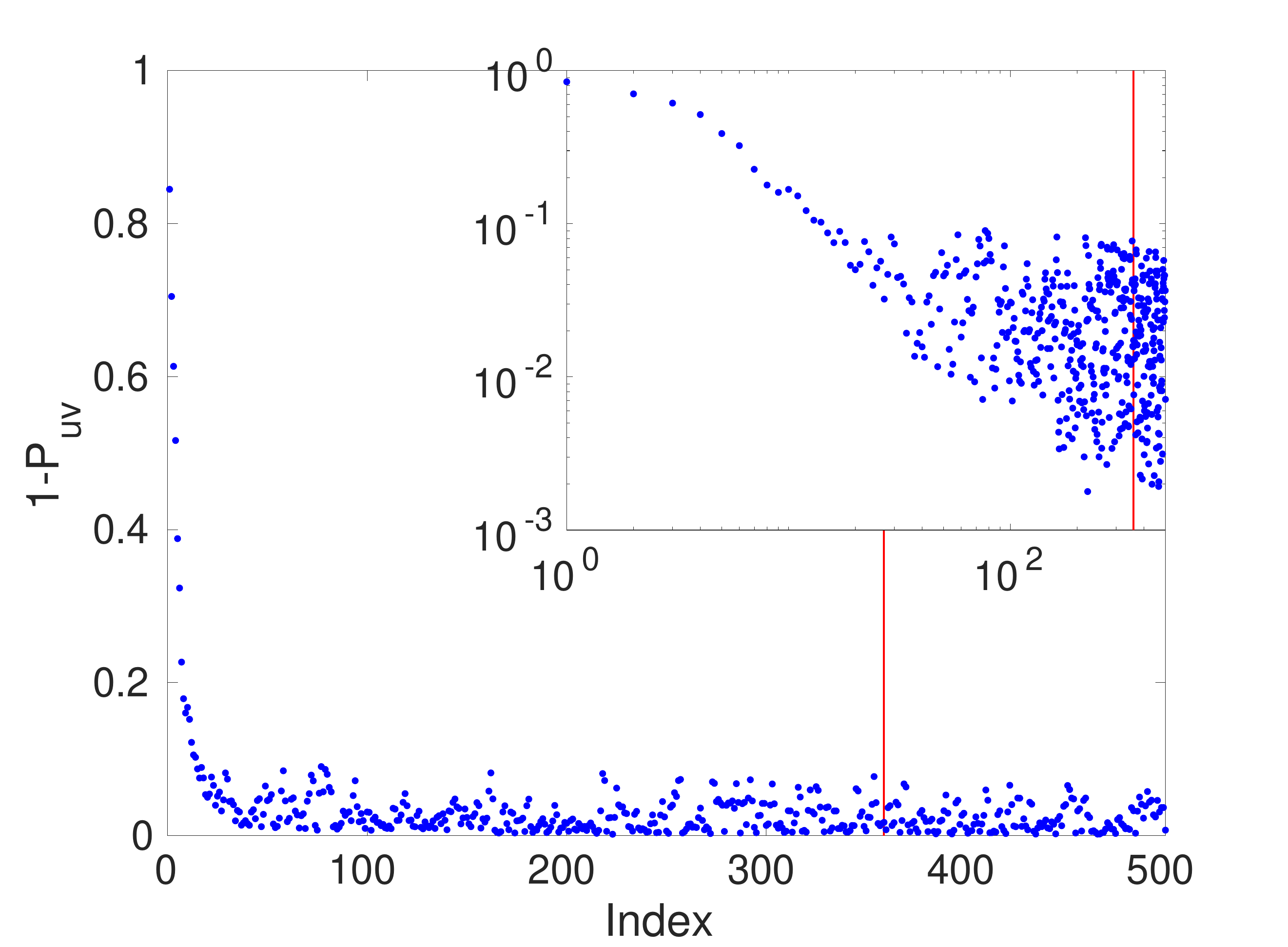}}
		\subfigure[]{\label{fig.s7_2} 
			\includegraphics[width=4cm]{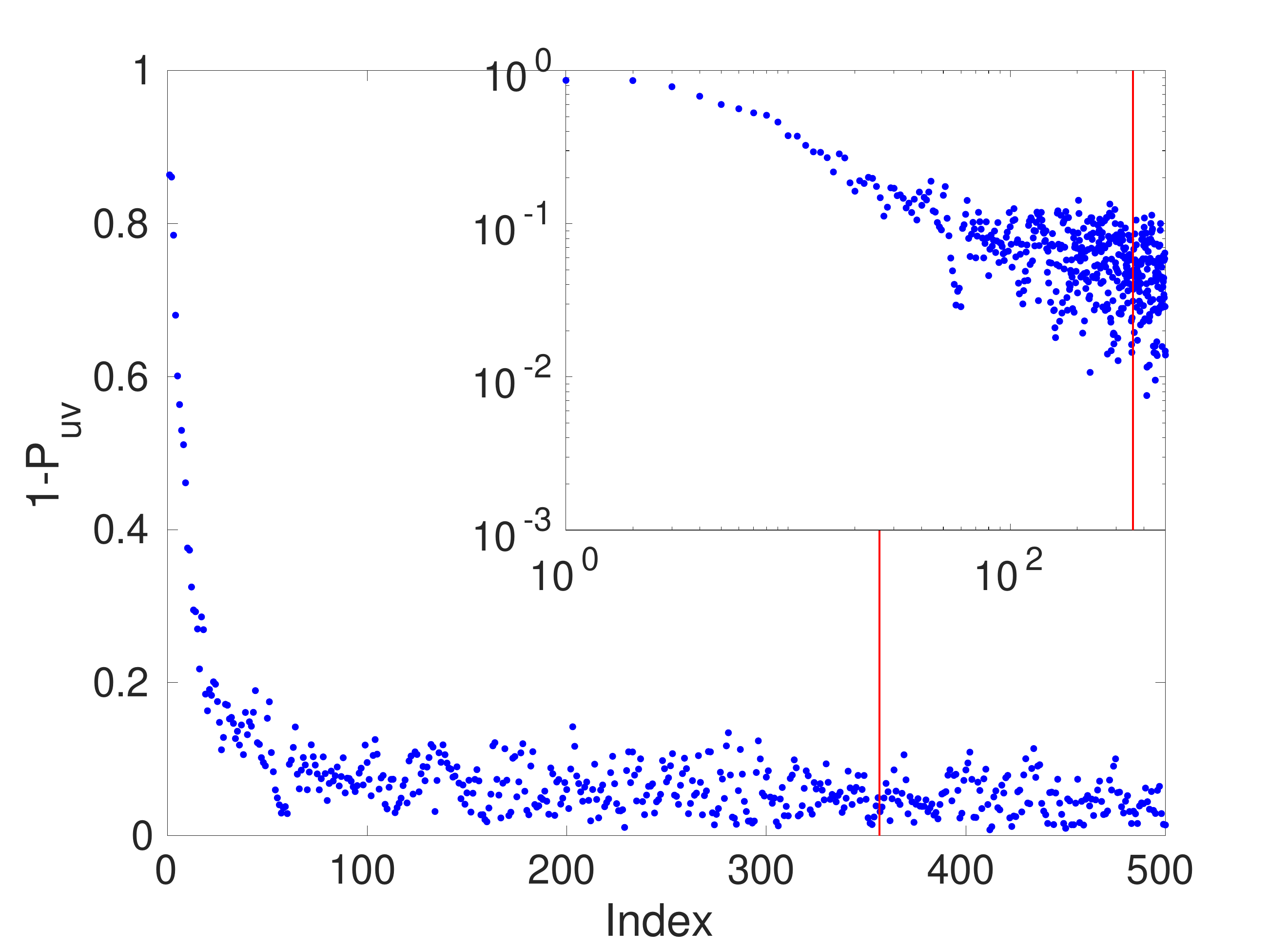}}
		\subfigure[]{\label{fig.s7_3} 
			\includegraphics[width=4cm]{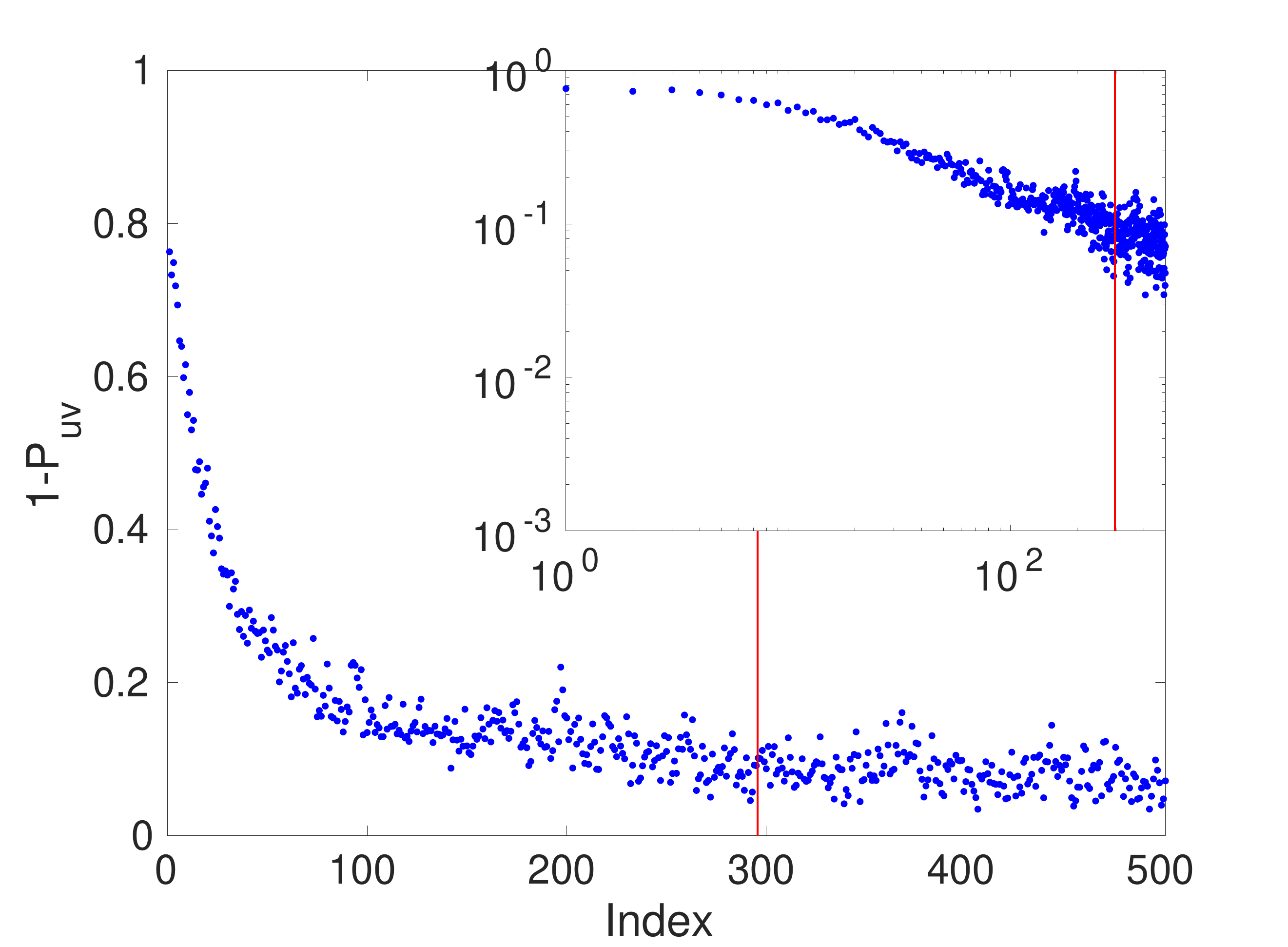}}
		\subfigure[]{\label{fig.s7_4} 
			\includegraphics[width=4cm]{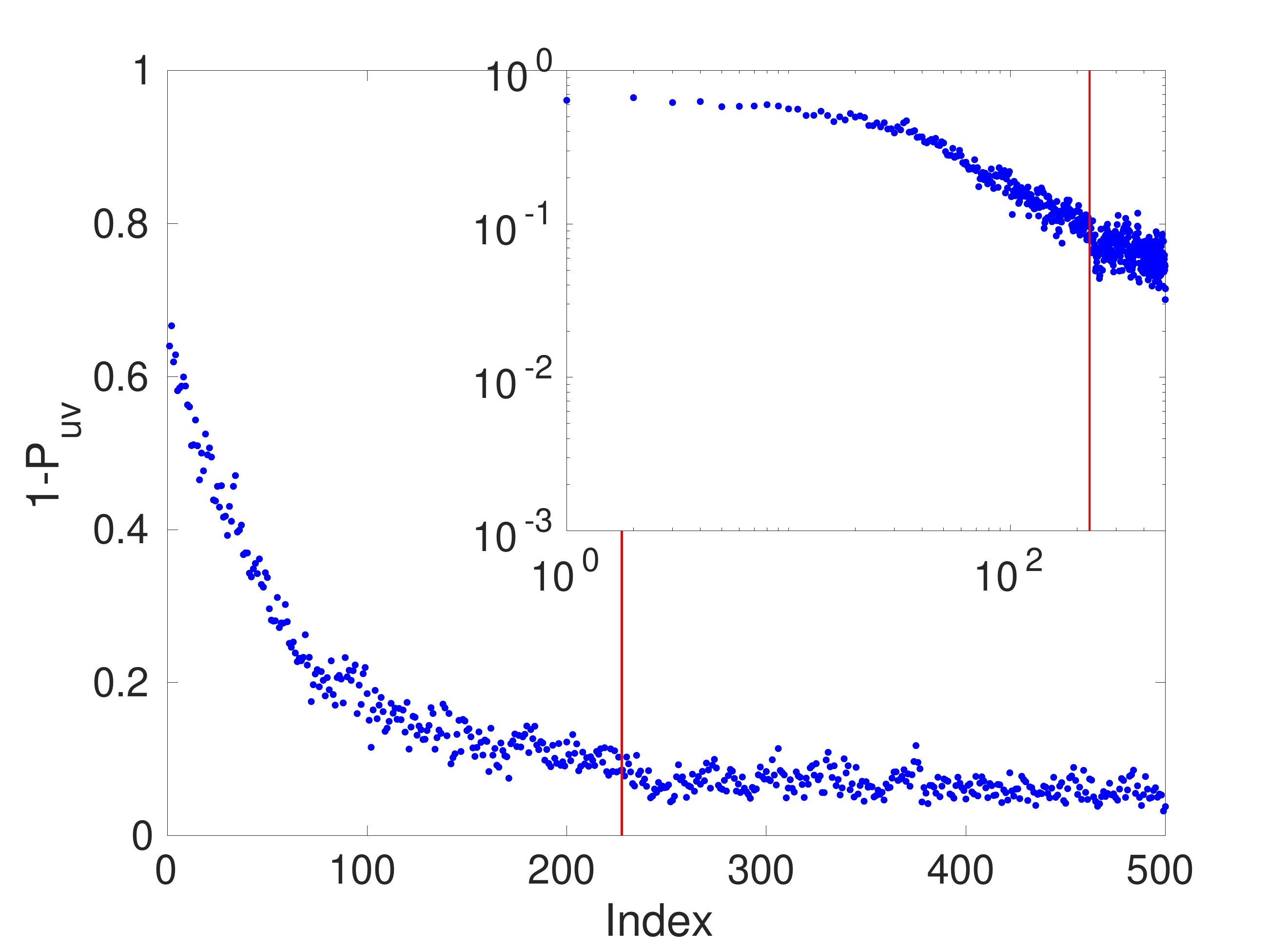}}\\
		\subfigure[]{\label{fig.s7_5} 
			\includegraphics[width=4cm]{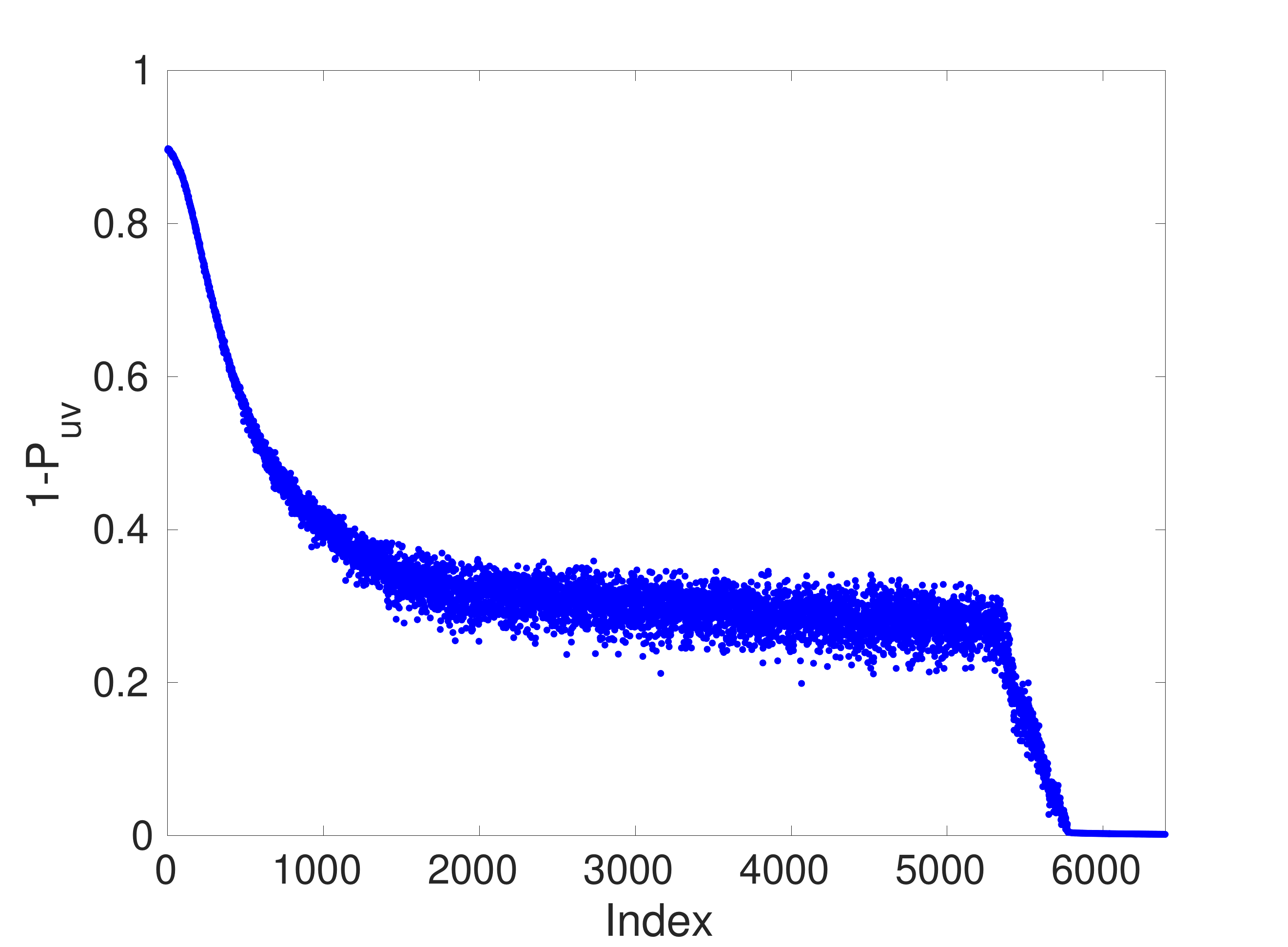}}
		\subfigure[]{\label{fig.s7_6} 
			\includegraphics[width=4cm]{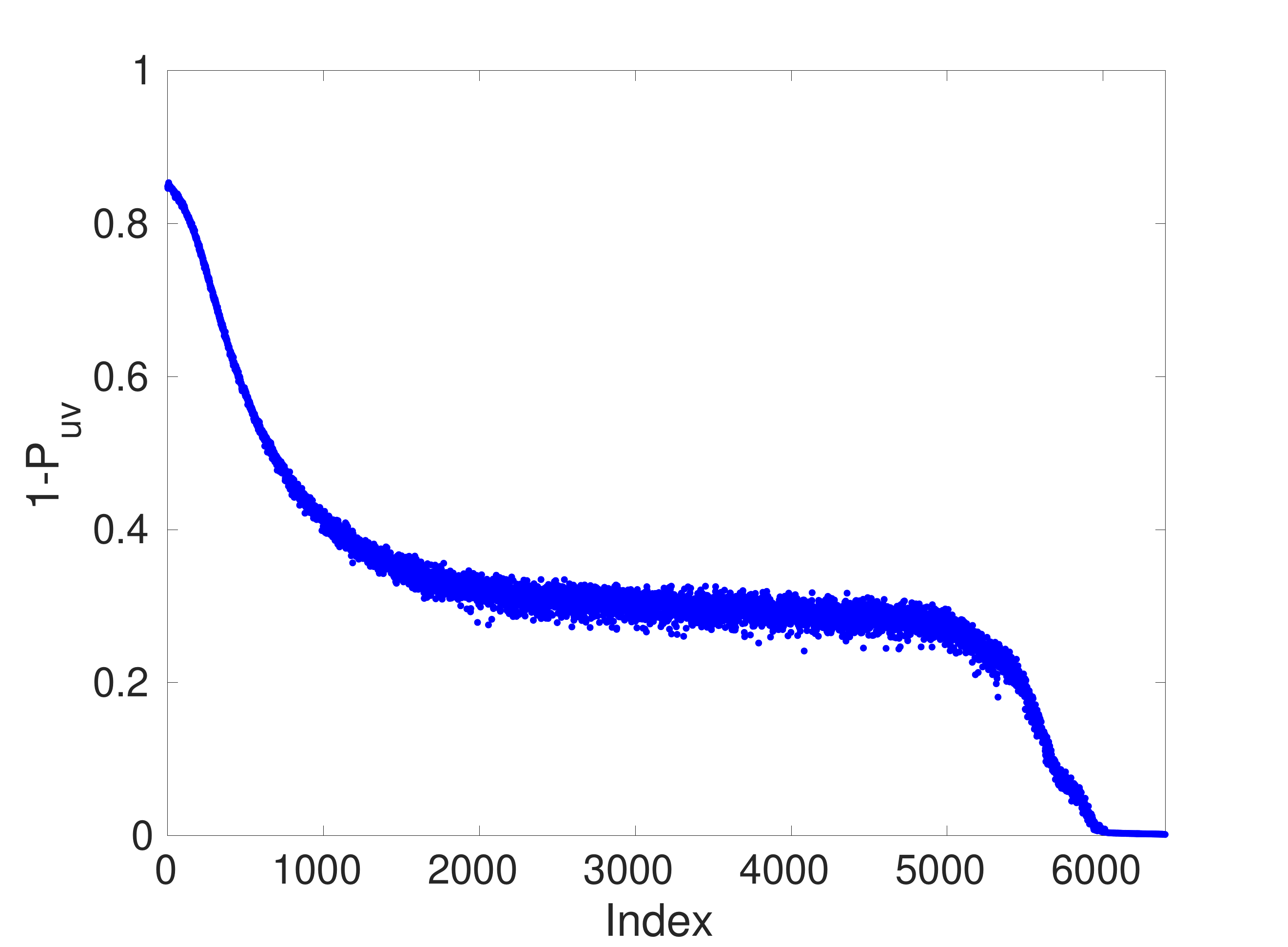}}
		\subfigure[]{\label{fig.s7_7} 
			\includegraphics[width=4cm]{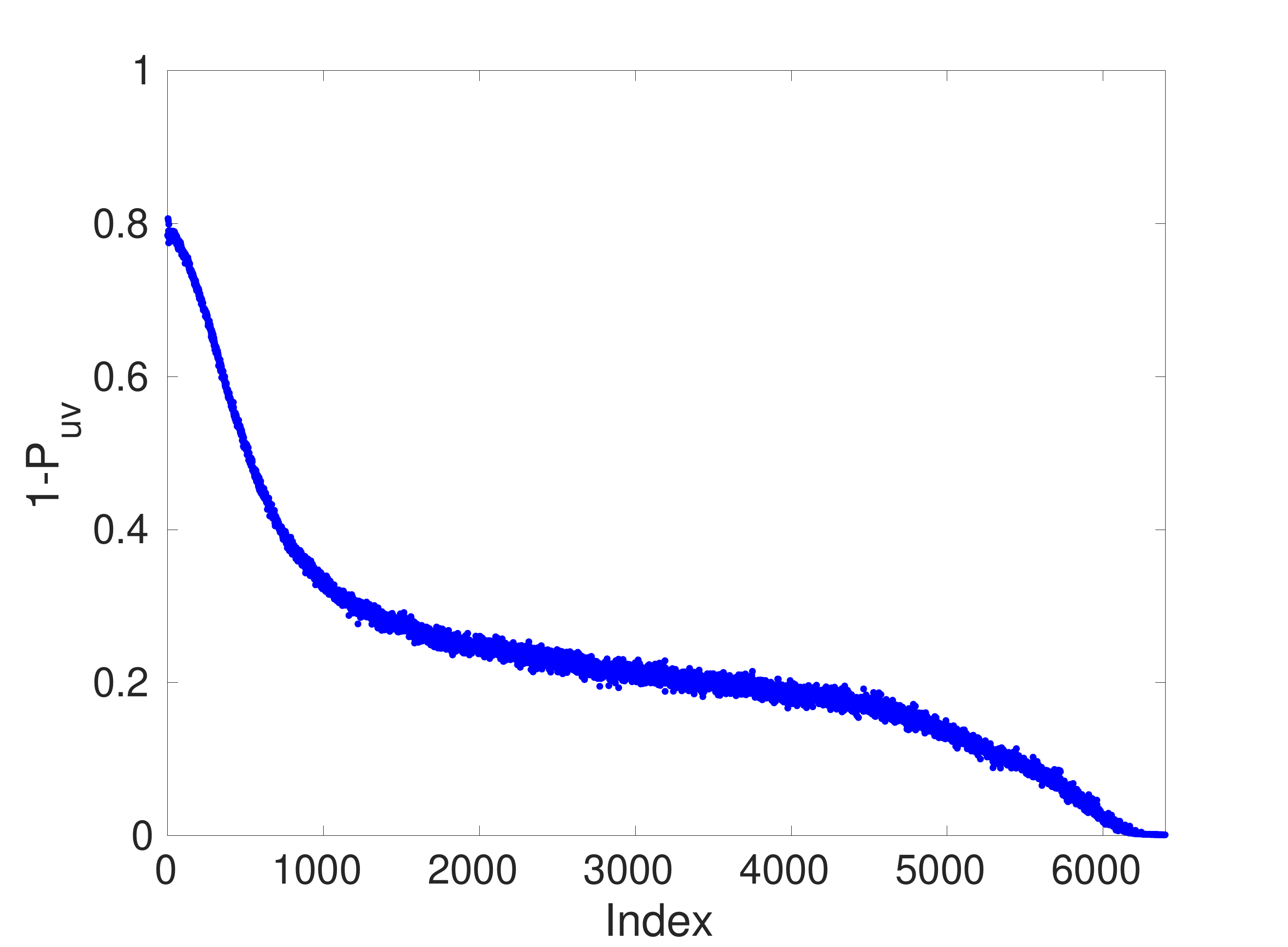}}
		\subfigure[]{\label{fig.s7_8} 
			\includegraphics[width=4cm]{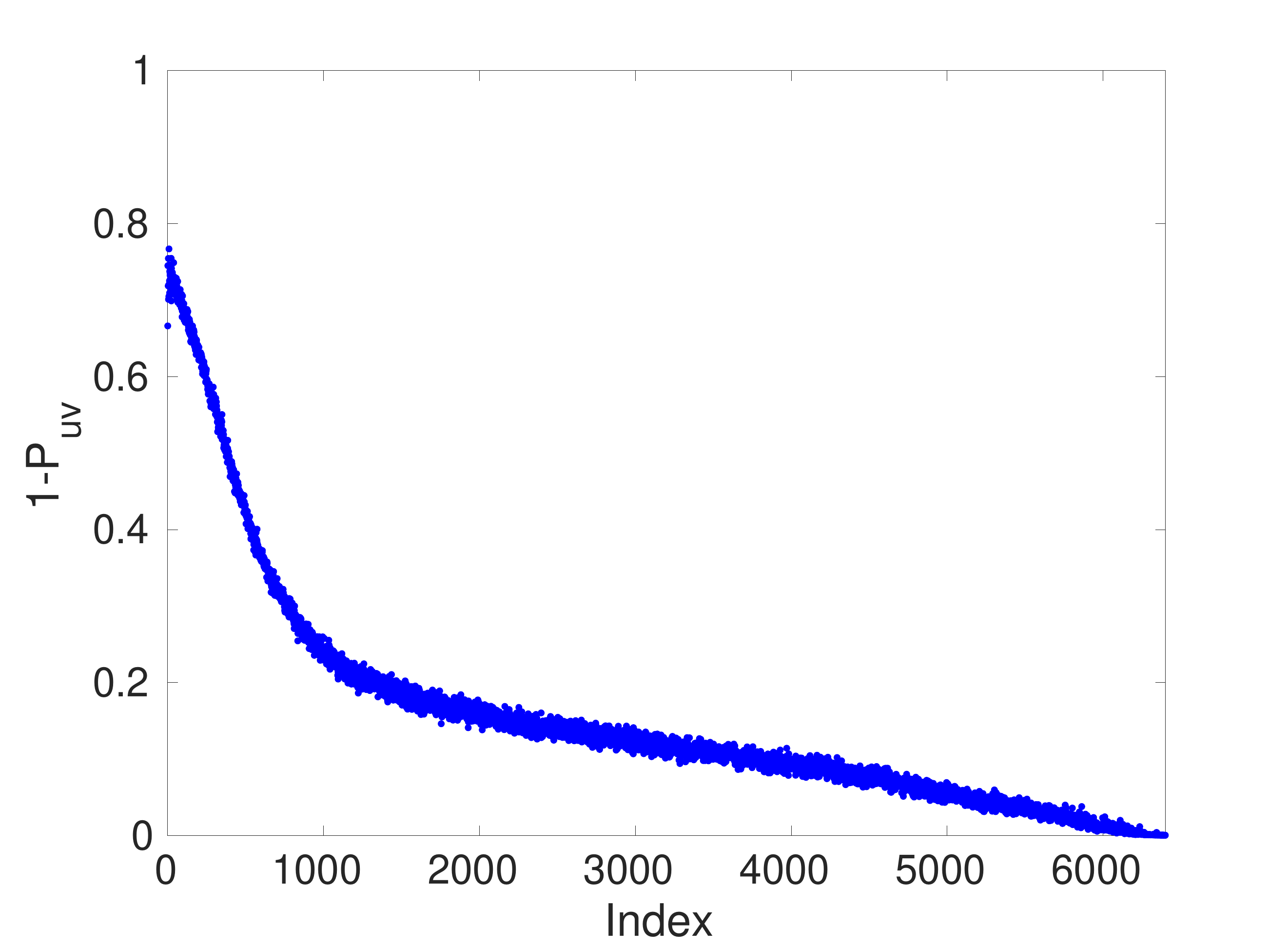}}
		\caption{$1-P_{uv}$ eq.~(\ref{puv}), in the weak (upper panel) and strong (lower panel) coupling limit for different disorder strength $V =0.5, 1, 2$, and $3$ from left to right.  
			Eigenfunctions corresponding to the lowest eigenvalues (whose energy is closest to $E = 0$), are almost identical and therefore $1-P_{uv} \approx 1$ both in the weak and in the strong coupling limit. However, in the weak coupling limit, there are very few eigenstates with such a strong correlation, a number much smaller than those in the Debye window (vertical red line). The number of strongly correlated eigenstates increases with disorder. By contrast, in the strong coupling limit, the coupling between $u_n$ and $v_n$ is strong close to the Fermi energy for any disorder and decreases rather slowly for higher energies. That could explain why disorder enhances superconductivity only in the weak coupling limit}\label{Fig.7}
	\end{center}
\end{figure}

In the week coupling limit, see Fig.~\ref{Fig.7}, we observe that only for a few states near $E = 0$, a number much less than the total number of states contained in the Debye window, $P_{uv}$ is close to $0$, while for the rest, $P_{uv} \approx 1$. Interestingly, as disorder increases, the number of strongly coupled eigenstates $P_{uv} \approx 0$ increases as well. Taking into account that $\Delta(r_i)$ is also defined through the overlap of $u_n(r_i)$ and $v_n(r_i)$, it is not surprising the previous result that disorder enhances the spatial average of $\Delta(r_i)$. Effectively, as disorder increases, more eigenstates contribute to the formation of the order parameter which likely help its enhancement. More quantitatively, about $100$ states are strongly coupled when disorder $V = 2$ and $3$, see Figs.~\ref{fig.s7_3} and \ref{fig.s7_4}, while such strong correlation is restricted to no more than $10$ eigenvectors for $V=0.5$.

Results in the strong coupling limit are quite different. As is observed in Fig.~\ref{Fig.7}, a majority of states are strongly coupled $(P_{uv}\ll 1)$. The coupling between $u_n(r)$ and $v_n(r)$ diminishes as the energy increases. Unlike the weak coupling limit, it seems the coupling between $u_n(r)$ and $v_n(r)$ is weaker as disorder increases which would explain why the order parameter decreases with disorder. We therefore expect that, in this limit, disorder always suppresses superconducting properties. Eigenstates near the edge of the spectrum are localized in space so they cannot be strongly coupled and $P_{uv} \approx 1$ in both the weak and the strong coupling.

\section{Level statistics of the Bogoliubov de-Gennes spectrum}
The study of spectral correlations is a powerful tool to characterize the quantum dynamics of many-body systems. It is specially suited to detect metal-insulator transitions induced by disorder. In a disordered metal, level statistics agrees with the prediction of random matrix theory (RMT), also called Wigner-Dyson statistics, characterized by level repulsion for short range correlations and level rigidity for correlations involving more distant eigenvalues. By contrast, in an Anderson insulator, spectral correlation are given by Poisson statistics, characterized by the absence of correlations among eigenvalues at all scales. At the Anderson transition, characterized by multifractal eigenstates, level statistics \cite{shapiro1993,altshuler1988repulsion,varga2000} have also distinct features: scale invariance, level repulsion for short range correlators as in a metal, and substantial weakening of level rigidity as in an insulator.

\begin{figure}[htbp]
	\begin{center}
		\subfigure[]{\label{fig.s8_1} 
			\includegraphics[width=5.5cm]{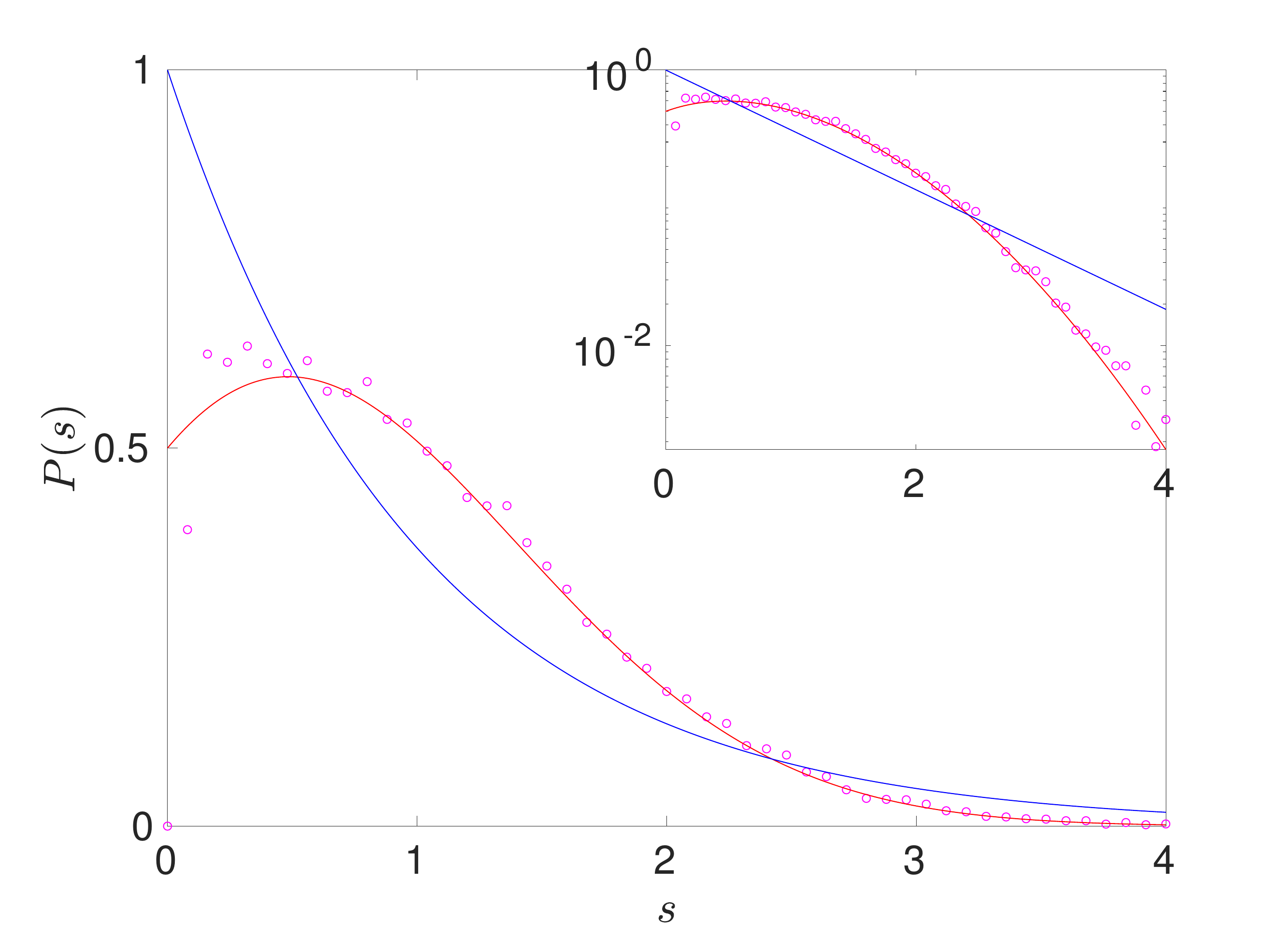}}
		\subfigure[]{\label{fig.s8_2} 
			\includegraphics[width=5.5cm]{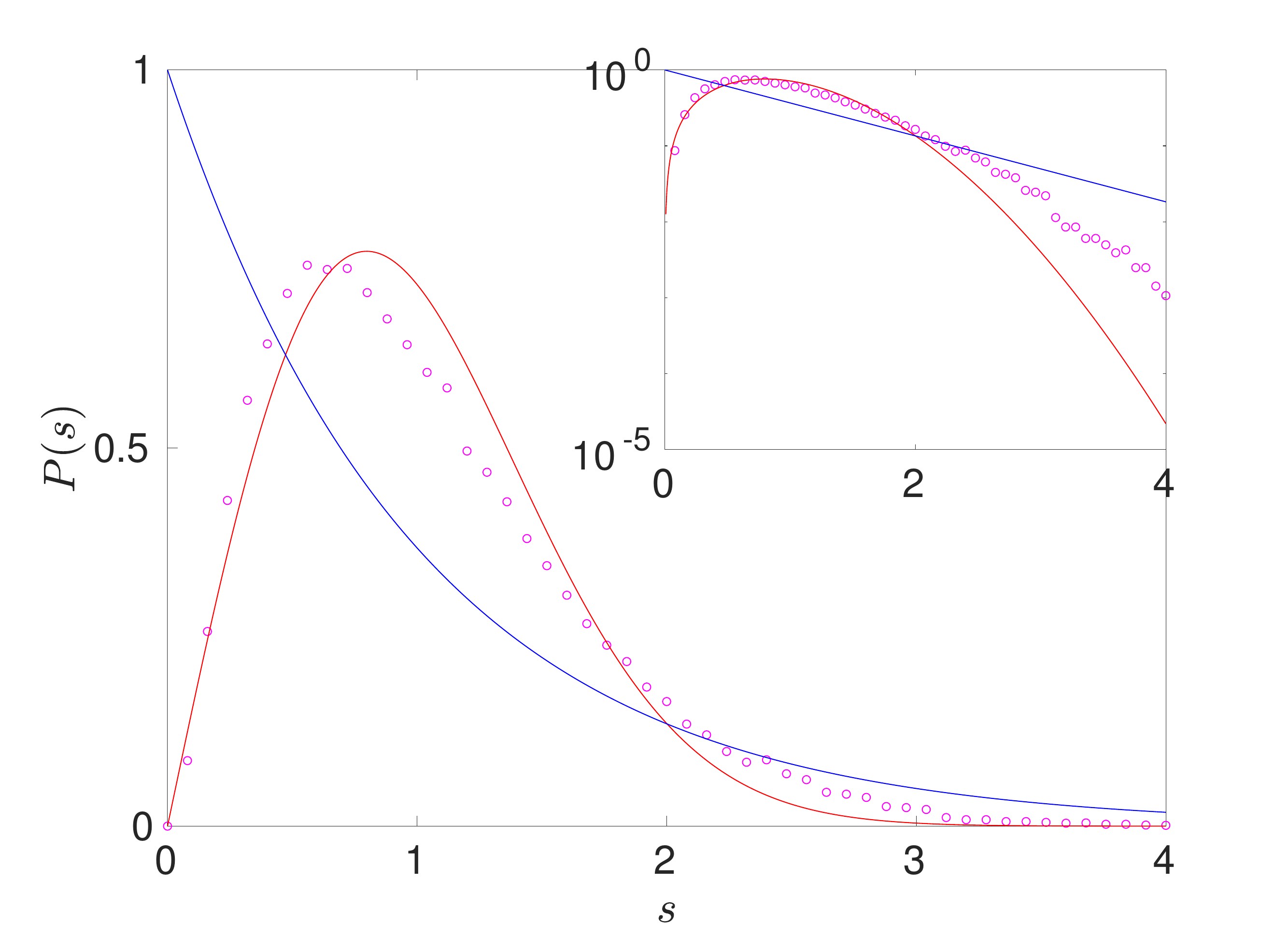}}
		\subfigure[]{\label{fig.s8_3} 
			\includegraphics[width=5.5cm]{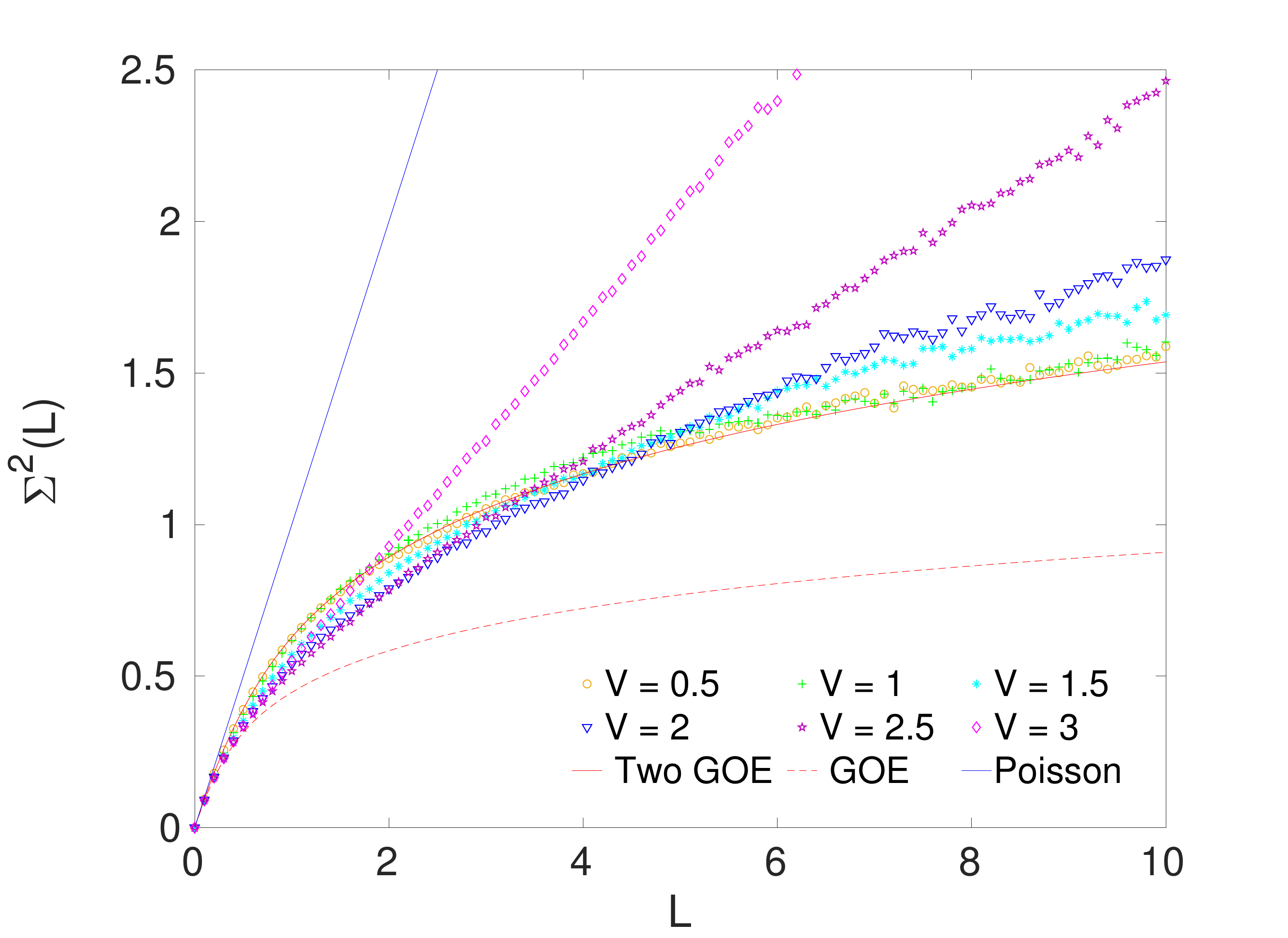}}
		\caption{Level spacing distribution $P(s)$ and number variance $\Sigma^2(L)$ in the weak coupling limit $U=-1$, $\omega_D =0.15$ (in units of $t$). 
		Only eigenvalues within the Debye energy window are taken into consideration.\subref{fig.s8_1}: weak disorder $V = 0.5$.  We find excellent agreement with the prediction for two superimposed spectra with Wigner-Dyson statistics. \subref{fig.s8_2}: strong disorder $V = 2$. Degeneracy is lifted, level repulsion is robust, and the decay for larger spacing is exponential. These spectral features are shared with systems at the metal-insulator transition \cite{shapiro1993,varga2000,wang2009}. \subref{fig.s8_3}: number variance for different disorder strength. Results are consistent with those of $P(s)$. For weak disorder, the number variance agrees well with that corresponding to the superposition of two spectra each following Wigner-Dyson statistics. For stronger disorder, it becomes linear but with a slope less than one which is consistent with the prediction for system close to an Anderson metal-insulator transition.} \label{Fig.8}
	\end{center}
\end{figure}

In this section we carry out an analysis of level statistics for the eigenvalues of the BdG equations. We restrict ourselves to the spectral region inside the Debye energy window since our main interest is to characterize the dynamics of the superconducting state. Before we proceed, it is important to note that the BdG equations have eigenvalues coming from eigenvectors $u_n(r)$, $v_n(r)$ representing the Cooper's pair. In the limit of no disorder, it is easy to see from the structure of the Hamiltonian eq.~(\ref{eq.1}), that the eigenvalues of $u_n(r)$ and $v_n(r)$ are two-fold degenerate. By turning on disorder, this degeneracy is lifted but for sufficiently weak disorder there is almost no mixing with neighboring eigenvalues so that effectively the full spectrum is the superposition of two spectra corresponding to weakly disordered metals. For sufficiently strong disorder, neighboring eigenvalues get mixed and the spectrum is no longer a superposition of two independent spectra. 

Results depicted in Fig.~\ref{Fig.8} fully confirm this picture. In the weak-disorder, weak-coupling limit, spectral correlation, both short and long range, agree well with the theoretical prediction for the superposition of two spectra with Wigner-Dyson statistics. The prediction for the level spacing distribution, namely, the probability of having two consecutive eigenvalues at a distance $s$ in units of the mean level spacing, is \cite{guhr1998}  $P(s) = \frac{\pi}{16}s(1-{\rm erf}(\sqrt{\pi}s/4))\exp(-\pi s^2/16)+\frac{1}{2}\exp(-\pi s^2/8)$, where ${\rm erf}(s)$ is the error function. Likewise, the number variance $\Sigma^2(L)$, namely, the variance in the number of eigenvalues in a spectral window in units of the mean level spacing. For sufficiently large $L$,  $\Sigma^2(L) = \frac{4}{\pi^2}(\ln(\pi L) + \gamma + 1 - \frac{\pi^2}{8})$, where $\gamma$ is Euler's constant and $L$ is a spectral window containing, after unfolding, $L$ eigenvalues on average. 

\begin{figure}[htbp]
	\begin{center}
		\subfigure[]{\label{fig.s9_1} 
			\includegraphics[width=5cm]{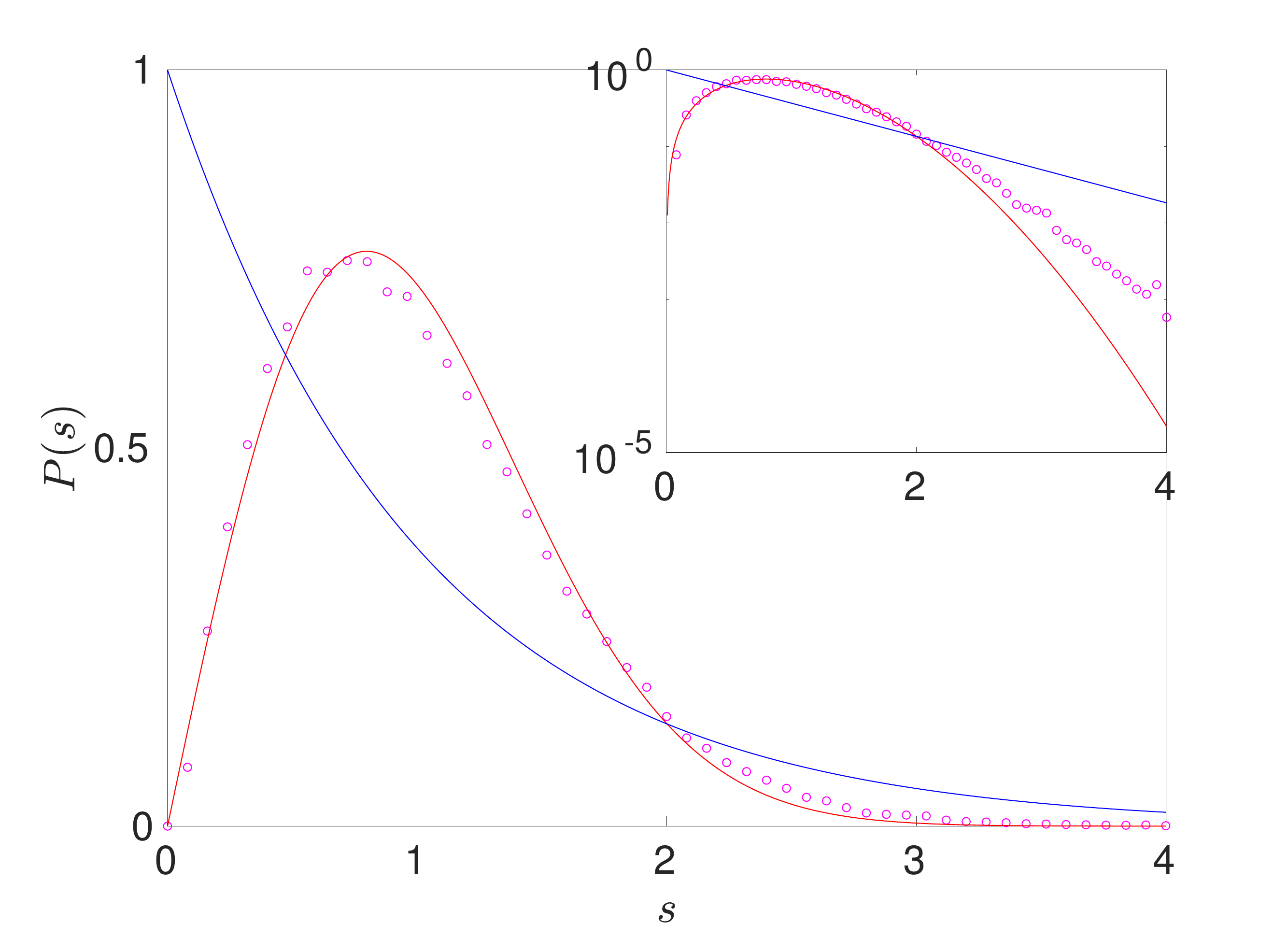}}
		\subfigure[]{\label{fig.s9_2} 
			\includegraphics[width=5cm]{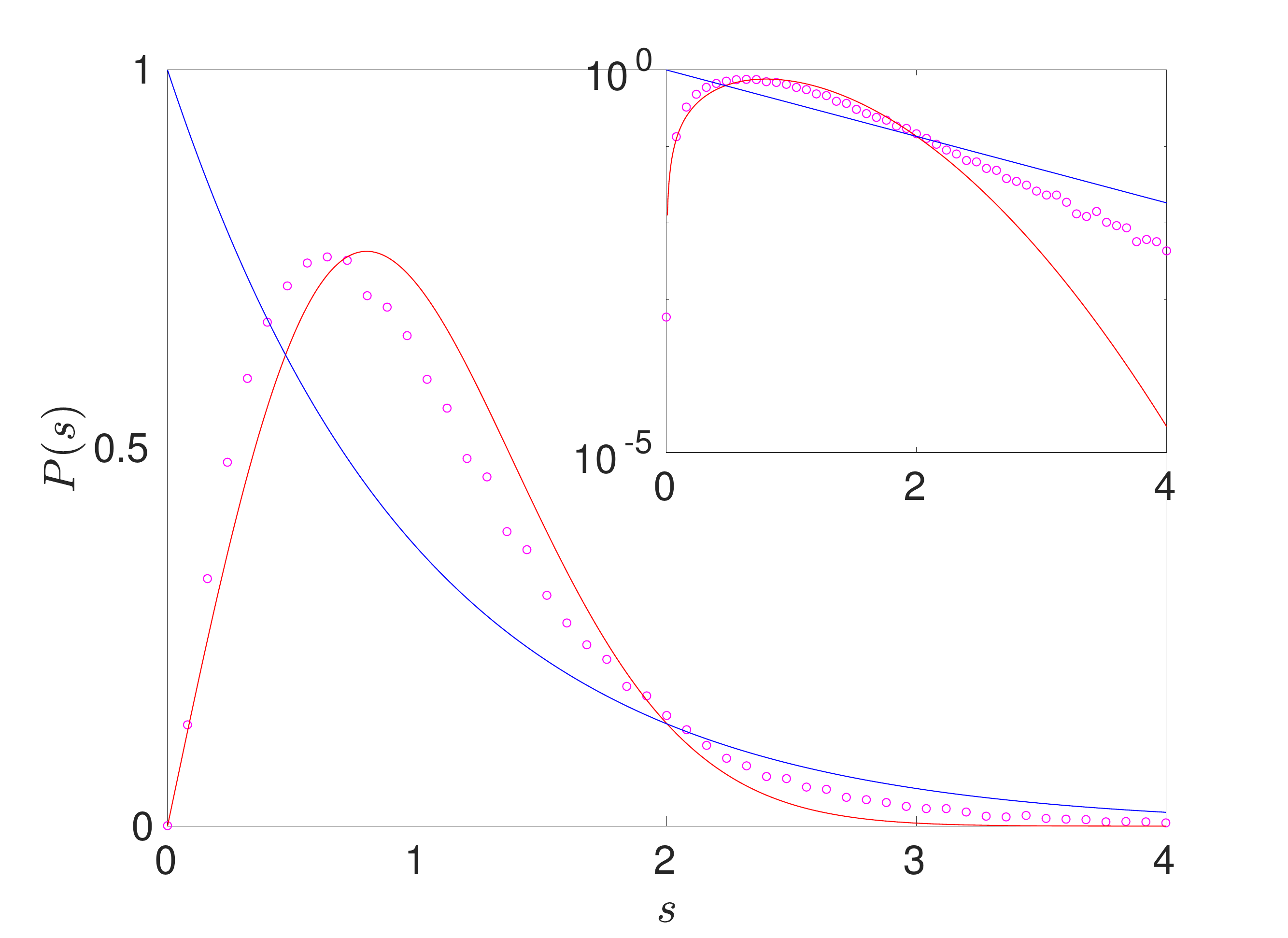}}
		\subfigure[]{\label{fig.s9_3} 
			\includegraphics[width=5cm]{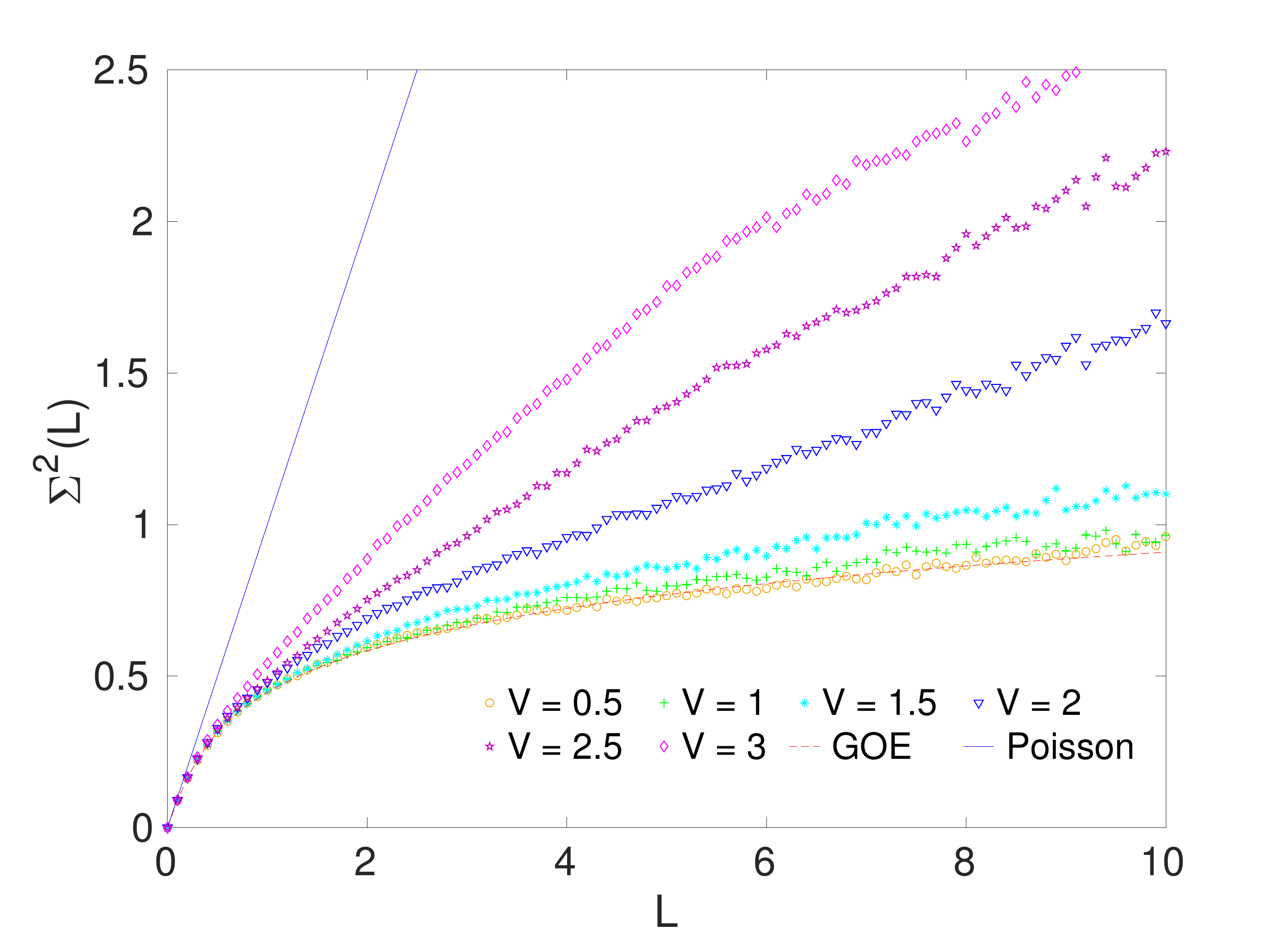}}
		\caption{The level spacing distribution $P(s)$ and the number variance $\Sigma^2(L)$ in the strong coupling limit, $\omega_D = \infty$, $U = -2$ and chemical potential $\mu = 0$. \subref{fig.s9_1}: $V = 0.5$, spectral correlations are close to Wigner-Dyson \subref{fig.s9_2}. $V = 3$: level repulsion is observed but the asymptotic decay is exponential as in an insulator.  \subref{fig.s9_3}:
		Only the first $200$ eigenvalues are taken into consideration for the number variance in order to mimic the Debye cut-off energy in the weak coupling limit. We observe that though spectral fluctuations deviate from Wigner-Dyson statistics, they are still very different from Poisson statistics which characterizes the insulating region. For $V \sim 1.5$, spectral correlations are similar to those of a system close to an Anderson metal-insulator transition \cite{shapiro1993}.}\label{Fig.9}
	\end{center}
\end{figure}

Our main motivation in this section is to determine whether in the range of disorder strength $V \leq 3$ that we investigate, the system is still in the metallic phase or a metal-insulator transition in the weak coupling limit occurs at $V = V_c <3$. Although only a calculation of the conductivity or other transport properties can conclusively answer this question, the study of level statistics is typically very reliable and technically much less demanding.

In Fig.~\ref{fig.s8_2}, we show results for the level spacing distribution $P(s)$ for $V = 2$ where we expect that the spectrum is sufficiently mixed. Indeed, level statistics are very close to those describing the metal-insulator transition in disordered and quantum chaotic systems: clear level repulsion is observed in the $P(s)$ for $s \ll 1$, while the decay for larger $s$ is exponential as in Poisson statistics but with a different decay exponent. Similarly, the number variance is linear but with a slope smaller than the prediction of Poisson statistics typical of an insulator. This is another indication that multifractality may play a role as this intermediate level statistics has been observed  in different type of systems \cite{shapiro1993,varga2000,altshuler1988repulsion,martin2008,wang2009} with multifractal eigenstates. 

Finally, we also investigate spectral correlations in the strong coupling limit (see Fig.~\ref{Fig.9}). Because of the stronger interaction, there is stronger mixing. As a consequence, the full spectrum is never a superposition of two independent spectra and Wigner-Dyson statistics applies. Even for stronger disorder $V \leq 3$, we do not observe a transition to Poisson statistics. Level statistics has metallic features such as a clear level repulsion even for the strongest disorder. Indeed, it is qualitatively similar to that in the weak-coupling case. It seems to also describe a system close to the transition though a finite size scaling analysis would be necessary to confirm this point. 

\section{Robustness of phase coherence and superconductor-metal-insulator transition}

So far we have restricted our analysis to the amplitude of the order parameter. However, defining features of superconductivity as zero resistance or phase coherence are related to the phase of the order parameter. In this section, we study the latter by computing the superfluid stiffness including phase fluctuations to leading order. We will show that phase coherence likely holds for intermediate disorder strength $V \sim 1$ where the order parameter has multifractal-like features and its spatial average is enhanced. At the disorder strength in which the stiffness vanishes, the spectrum and eigenfunction has still metallic features which indicates that phase coherence is lost while the system is still in the metallic phase. This is suggestive of an intermediate dirty Bose metal phase in this system. 

Phase coherence, a defining feature of a superconductor, occurs if the superfluid stiffness is finite. In order to make an estimate of the disorder strength at which phase coherence is lost, we first use the two dimensional quantum XY model with the effective Hamiltonian \cite{ramakrishnan1989superconductivity,ghosal1998} $H_\theta = -(\kappa/8)\sum_{j}\dot{\theta}_j^2+D_s^0/4\sum_{ij}\cos(\theta_i-\theta_j)$, where $\kappa = dn/d\mu$ is the compressibility, and $D_s^0/\pi = \langle -k_x \rangle - \Lambda_{xx}(q_x=0,q_y\rightarrow0,\omega=0)$ is the mean field phase stiffness\cite{PhysRevB.47.7995}, $\langle -k_x \rangle$ is the kinetic energy and $\Lambda_{xx}$ is the transverse current-current correlation function. Phase fluctuations are considered by using the self-consistent Gaussian approximation \cite{Ghosal2001,PhysRevB.25.1600,PhysRevLett.56.2303} leading to a renormalized superfluid stiffness,
\begin{equation}
	D_s = D_s^0 \exp(-\langle \theta_{ij}^2 \rangle_0/2) \label{eq.11}
\end{equation}
where $\langle \theta_{ij}^2 \rangle_0$ is the mean square fluctuation of the nearest neighbor phase difference $\langle \theta_{ij}^2 \rangle_0 = \frac{2}{N\xi} \sum_{q}\sqrt{\left(\frac{\epsilon_q}{D_s\kappa}\right)}$, and $\epsilon_q = 2[2-\cos(q_x)-\cos(q_y)]$. Therefore, it remains to compute $D_s^0$, the coherence length $\xi$ and the compressibility $\kappa$.

\begin{figure}[htbp]
	\begin{center}
		\subfigure[]{\label{fig.s10_1} 
			\includegraphics[width=5.5cm]{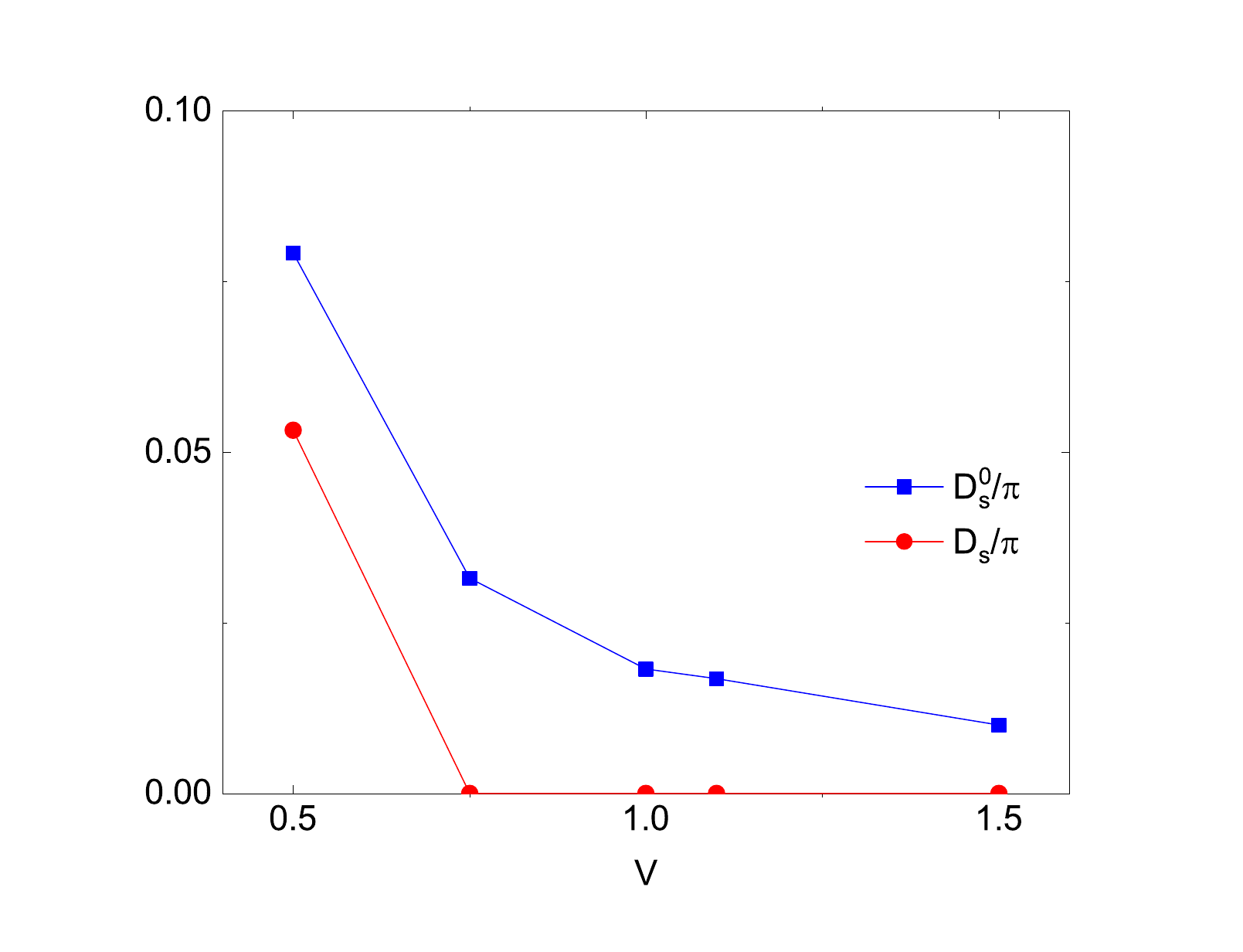}}
		\subfigure[]{\label{fig.s10_2} 
			\includegraphics[width=5.5cm]{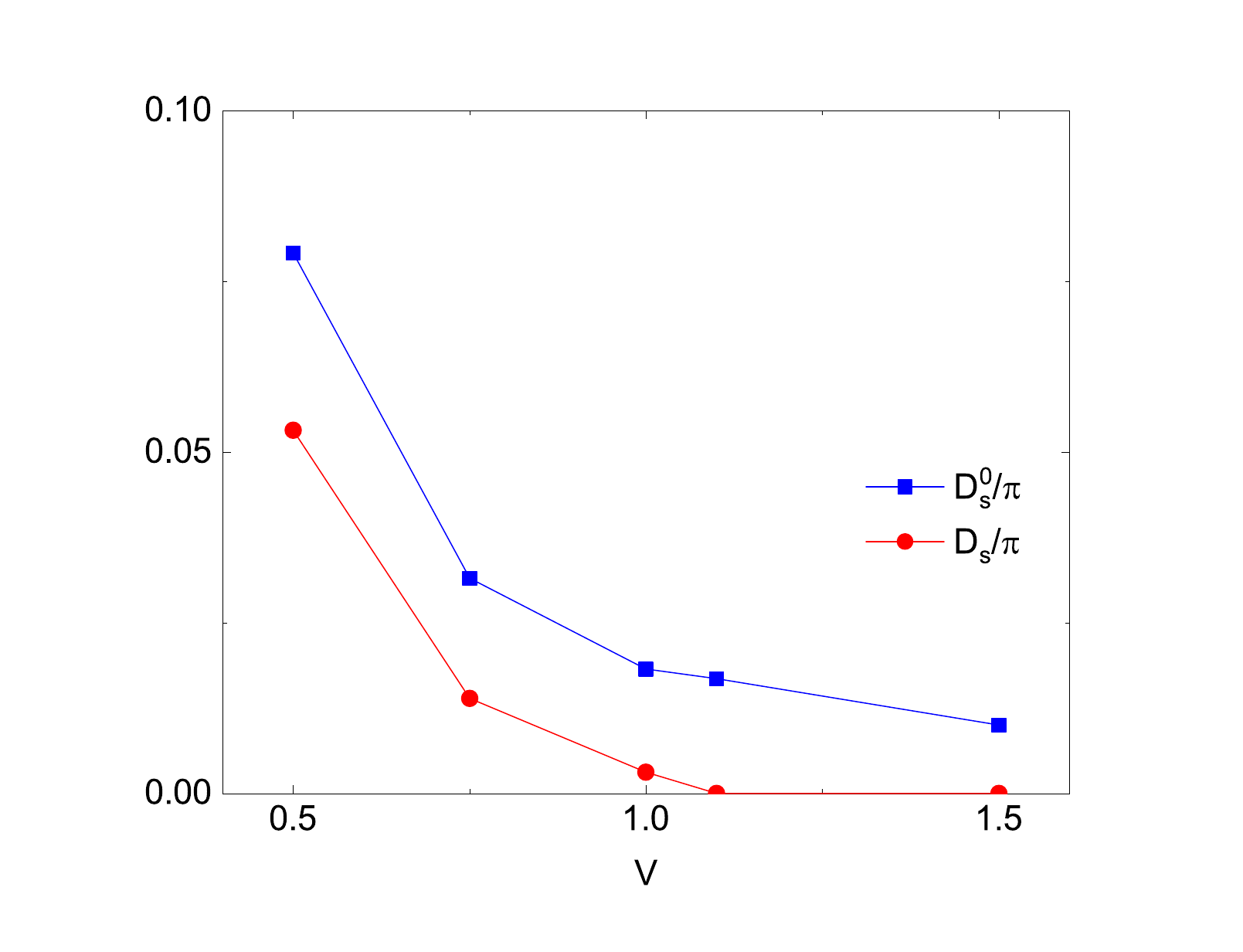}}
		\subfigure[]{\label{fig.s10_3} 
			\includegraphics[width=5.5cm]{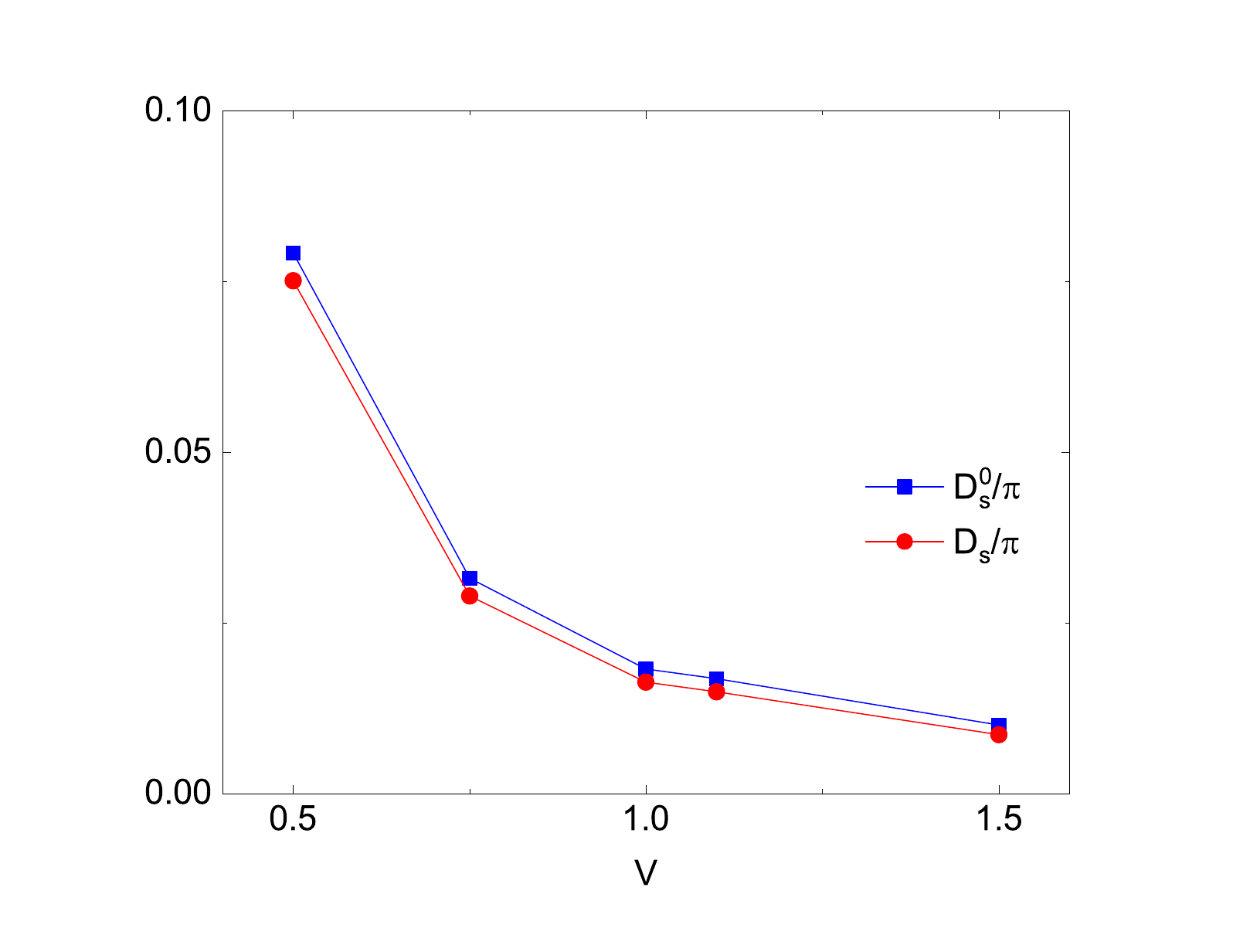}}
		\caption{The mean field superfluid stiffness $D_s^0/\pi$ and the full superfluid stiffness $D_s/\pi$ (\ref{eq.11}) in units of $t$, including phase fluctuations, as a function of disorder. 
		The chemical potential is $\mu = 0$, the system size is $80\times80$, $\omega_D = 0.15$ and $U = -1$.  As is observed, results are rather sensitive to the coherence length. \subref{fig.s10_1}: coherence length is that of Fig.~\ref{Fig.2}. \subref{fig.s10_2}: using previous choices in the literature \cite{Ghosal2001,ghosal1998} which in our case corresponds to	
		 $\xi \approx 15.727$, the coherence length for $V = 0.5$. \subref{fig.s10_3}: coherence length $\xi_0 \approx 100$ is that of the clean limit which leads to $D_s \approx D_s^0$.}\label{Fig.10}
	\end{center}
\end{figure}

We focus on the region of intermediate disorder as our main aim to find out whether superconductivity, as defined by a finite $D_s$, is robust to the strong spatial fluctuations of the order parameter amplitude. 
The answer is mostly affirmative. As is observed in Fig.~\ref{Fig.10}, the transition is located around $V = V_c \approx 1$ provided that, following  Ref.\cite{ghosal1998,Ghosal2001}, the chosen coherence length is the largest one that it is still much smaller than the system size. 

As shown previously, for $V \sim 1$, the averaged order parameter $\bar \Delta$ is enhanced by disorder, its spatial distribution is log-normal and the singularity spectrum $f(\alpha)$ is parabolic. All these are typical features in systems where multifractality plays a role. However, it is important to stress that $D_s$ is very sensitive to the choice of coherence length. Theoretically, there is some ambiguity in the derivation of eq.~(\ref{eq.11}). It is assumed that the superconducting coherence length does not depend on disorder. 
According to Refs.\cite{ghosal1998,Ghosal2001}, our choice, at least in the strong coupling limit, is consistent with Monte Carlo simulations. Another point worth to mention is that the numerical calculation is rather demanding due to the large lattice size $80 \times 80$. For this reason, only one disorder realization is considered. Results for smaller sizes indicate that ensemble fluctuations are very small and therefore one disorder is in principle enough.

In summary, superconductivity, even in the two dimension limit, is robust to strong quantum coherence effects induced by a relatively weak disorder. We note that the results for larger lattices may shift the transition to a slightly stronger disorder because finite size effects enhance phase fluctuations. 

\section{Conclusion and outlook}
We have studied two-dimensional, weakly-disordered, weakly-coupled superconductors by solving numerically the Bogoliubov de-Gennes equations. The effect of small phase fluctuations has also been taken into account. We have found that, unlike strongly coupled superconductors, the spatial distribution of the order parameter amplitude is log-normal and the average gap and order parameter amplitude increases with disorder. These results are in rough agreement with analytical findings based on a simpler, not fully self-consistent BCS formalism \cite{Mayoh2015}. The analysis of the $f(\alpha)$ spectrum reveals that the probability measure related to the order parameter is closer to parabolic which is the prediction for weakly multifractal eigenstates. Level statistics for intermediate disorder are consistent with those of a critical system at the Anderson metal-insulator transition. Therefore our results suggest that enhancement of superconductivity and some approximated form of multifractality may be generic features of two dimensional weakly-disordered, weakly-coupled superconductors. A natural question to ask is why multifractality may play any role as this concept is associated to systems with no scales and we expect that the superconducting coherence length is still a typical length in our problem. Indeed, in the strong coupling limit, this length is clearly observed in the spatial dependence of the gap. It is possible this would also be the case in the weak coupling limit though it would be necessary much longer lattice sizes to observe it. A finite size scaling analysis may shed light on this issue. Even if it does, multifractality would still be relevant in a broad range of lengths which could impact transport and other properties of two dimensional superconducting materials.

The calculation of the superfluid density, computed including the effect of phase fluctuations, for intermediate disorder $V \sim 1$, reveals that phase coherence coexists with the intricate spatial distribution of the order parameter. The vanishing of the density occurs for larger disorder though the order parameter amplitude is still finite and level statistics are far from those of an insulator. This is reminiscent of a dirty Bose metal phase though further research, likely involving the study of transport properties like the conductivity, is required to test the reach of this similarity.  

It would be interesting to investigate finite temperature and magnetic field effects that could shed light on long standing problems in low dimensional superconductivity such as the conditions for enhancement of the critical temperature, the nature of the dirty Bose metal or the inhomogeneous Kosterlitz-Thouless transition. In the latter, the spatial distribution of vortices, likely very inhomogeneous, can modify qualitatively  the transition. We are currently investigating some of these problems.
\acknowledgments 
A.M.G.G. acknowledges financial support from a Shanghai talent program and from the National Natural Science Foundation of China (NSFC) (Grant number 11874259).
\bibliographystyle{unsrt}
\bibliography{library} 
\end{document}